%% file: prx_main.tex
\documentclass[aps, prx, twocolumn, superscriptaddress, floatfix, nofootinbib, longbibliography]{revtex4-2}
\usepackage{graphicx, amsmath, amsfonts, amssymb, amsthm, xr}
\usepackage{epsfig,amsmath,amssymb,color,dsfont,upgreek,physics}
\usepackage{mathrsfs}
\usepackage{braket}
\usepackage{mathtools}
\usepackage{comment}
\usepackage{bbold}
\usepackage{float}
\usepackage[caption=false]{subfig}
\usepackage[bookmarks=true,colorlinks,linkcolor=OrangeRed,urlcolor=NavyBlue,citecolor=RoyalBlue]{hyperref}
\usepackage[capitalize]{cleveref}
\usepackage[dvipsnames]{xcolor}
\usepackage{orcidlink}

\usepackage{standalone}
\usepackage{tikz-3dplot}
\usepackage{tikz}
\usetikzlibrary{3d,perspective,arrows.meta}

\definecolor{color_upper}{HTML}{1E88E5}
\definecolor{color_lower}{HTML}{105EA2}
\definecolor{color_boundary}{HTML}{D81B1B}
\definecolor{color_sub}{HTML}{FFC107}
\definecolor{color_geo}{HTML}{004D40}

\newcommand{\be}{\begin{equation}}
\newcommand{\ee}{\end{equation}}
\newcommand{\ba}{\begin{eqnarray}}
\newcommand{\ea}{\end{eqnarray}}

\usepackage[dvipsnames]{xcolor}

\DeclareMathOperator{\arctanh}{arctanh}
\newcommand{\vertiii}{\vert\kern-0.25ex\vert\kern-0.25ex\vert}

\newcommand{\CR}{\text{cross-ratio}}

\begin{document}

\title{Dynamical Entanglement Phase Transitions in Holographic CFTs}


\author{Lukas Ebner${}^{\orcidlink{0009-0005-5186-9042}}$}
\affiliation{Department of Physics and Arnold Sommerfeld Center for Theoretical Physics (ASC), Ludwig Maximilian University of Munich, 80333 M\"unchen, Germany}
\affiliation{Max Planck Institute of Quantum Optics, 85748 Garching, Germany}
\affiliation{Munich Center for Quantum Science and Technology (MCQST), 80799 M\"unchen, Germany}

\author{Jad C.~Halimeh${}^{\orcidlink{0000-0002-0659-7990}}$}
\affiliation{Department of Physics and Arnold Sommerfeld Center for Theoretical Physics (ASC), Ludwig Maximilian University of Munich, 80333 M\"unchen, Germany}
\affiliation{Max Planck Institute of Quantum Optics, 85748 Garching, Germany}
\affiliation{Munich Center for Quantum Science and Technology (MCQST), 80799 M\"unchen, Germany}
\affiliation{Department of Physics, College of Science, Kyung Hee University, Seoul 02447, Republic of Korea}

\author{David Horn${}^{\orcidlink{0000-0002-6288-4791}}$}
\affiliation{Institut f\"ur Theoretische Physik, Universit\"at Regensburg, Regensburg, D-93040, Germany}
\affiliation{Department of Physics, Duke University, Durham, 27708-0305, NC, USA}

\author{Joseph Dominicus Lap${}^{\orcidlink{0000-0002-6313-890X}}$}
\email{Joseph.Lap@epfl.ch}
\affiliation{Fields and Strings Laboratory, Institute of Physics, \'Ecole Polytechnique F\'ed\'erale de Lausanne
(EPFL), CH-1015 Lausanne, Switzerland}
\affiliation{Center for Quantum Science and Engineering, \'Ecole Polytechnique F\'ed\'erale de Lausanne
(EPFL), CH-1015 Lausanne, Switzerland}

\author{Jakob J.~Minar${}^{\orcidlink{0009-0008-5557-6919}}$}
\affiliation{Institut f\"ur Theoretische Physik, Universit\"at Regensburg, Regensburg, D-93040, Germany}

\author{Berndt M\"{u}ller${}^{\orcidlink{0000-0003-2468-3996}}$}
\affiliation{Department of Physics, Duke University, Durham, 27708-0305, NC, USA}

\author{Andreas Sch\"{a}fer${}^{\orcidlink{0000-0003-1171-0078}}$}
\affiliation{Institut f\"ur Theoretische Physik, Universit\"at Regensburg, Regensburg, D-93040, Germany}

\author{Clemens Seidl${}^{\orcidlink{0009-0006-9409-7553}}$}
\affiliation{Department of Physics and Arnold Sommerfeld Center for Theoretical Physics (ASC), Ludwig Maximilian University of Munich, 80333 M\"unchen, Germany}
\affiliation{Munich Center for Quantum Science and Technology (MCQST), 80799 M\"unchen, Germany}
\affiliation{Institut f\"ur Theoretische Physik, Universit\"at Regensburg, Regensburg, D-93040, Germany}

\date{\today}

\begin{abstract}
We study the time evolution of the entanglement structure of holographic conformal field theories after a local quench. Using the mutual information between two spatial intervals as a probe, we find that $1+1$-dimensional conformal field theories exhibit a rich pattern of dynamical phase transitions. In the large-central-charge limit, mutual information develops sharp non-analyticities at critical times, providing a concrete entanglement-based realization of dynamical quantum phase transitions. We find that the dynamics organize into six distinct phases of mutual information, each controlled by the dominance of a different conformal block, or equivalently, a different holographic geodesic configuration. This phase structure goes beyond the standard quasi-particle picture, explaining non-analytic features that are not captured by simple light-cone propagation from the quench points. We further identify a dynamical $D_4$ symmetry acting on the interval endpoints that controls the presence or absence of mutual information. The onset of mutual information is governed by the breaking of this symmetry to a $\mathbb{Z}_2 \times \mathbb{Z}_2$ subgroup, suggesting a symmetry-based characterization of non-equilibrium entanglement dynamics analogous to the role of symmetry in equilibrium critical phenomena. Finally, numerical studies of critical spin chains indicate that finite‑$c$ effects smooth out the sharp large‑$c$ transitions between different mutual‑information phases, while the transitions between phases with and without mutual information appear to remain non‑analytic. These results offer a unifying perspective on real-time entanglement dynamics and their critical features in conformal many-body systems.
\end{abstract}

\maketitle

\tableofcontents

\section{Introduction}
\input{Sections/1_Introduction}

\label{sec:section1}

\section{Review and Theoretical Background}
\label{sec:section2}
\input{Sections/2_Review}

\section{Classification of Phases in the UHP}
\label{sec:section3}
\input{Sections/3_Classification_of_Phases}

\section{Dynamical Phase Transitions in Mutual Information}
\label{sec:section4}
\input{Sections/4_Dynamical_Phase_Transitions}

\section{Finite Central Charge}
\label{sec:section5}
\input{Sections/5_Finite_c_Effects}

\section{Discussion}
\label{sec:section6}
\input{Sections/6_Discussion}

\begin{acknowledgments}
The authors gratefully acknowledge the scientific support and HPC
resources provided by the Erlangen National High Performance Computing
Center (NHR@FAU) of the Friedrich-Alexander-Universität Erlangen-Nürnberg
(FAU) under the NHR project b172da. NHR funding is provided by federal and
Bavarian state authorities.
D.H.~and B.M.~are supported by the National Science Foundation (Project PHY-2434506). B.M.~also acknowledges support by the U.S.~Department of Energy, Office of Science (Grant DE-FG02-05ER41367).
L.E.~and J.C.H.~acknowledge funding by the Max Planck Society, the Deutsche Forschungsgemeinschaft (DFG, German Research Foundation) under Germany’s Excellence Strategy – EXC-2111 – 390814868, and the European Research Council (ERC) under the European Union’s Horizon Europe research and innovation program (Grant Agreement No.~101165667)—ERC Starting Grant QuSiGauge. Views and opinions expressed are those of the author(s) only and do not necessarily reflect those of the European Union or the European Research Council Executive Agency. Neither the European Union nor the granting authority can be held responsible for them. This work is part of the Quantum Computing for High-Energy Physics (QC4HEP) working group.
A.S.~and C.S.~are supported by the DFG (German Research Foundation, grant 553079183).
C.S.~acknowledges financial support from the German Academic Scholarship Foundation.
This research was supported in part by grant NSF PHY-2309135 to the Kavli Institute for Theoretical Physics (KITP).
\end{acknowledgments}

\appendix
\section{Conformal Maps}
\label{app:maps}
\input{Sections/Appendix_A_Conformal_Maps}

\section{Phase Transition Inequalities}
\label{app:phases}
\input{Sections/Appendix_B_Phase_Transition_Inequalities}

\section{Group Structure of Phases}
\label{app:group structure}
\input{Sections/Appendix_C_Group_Structure_of_Phases}

\section{Lattice Details}
\label{app:lattice_details}
\input{Sections/Appendix_D_lattice_details}

\bibliography{refs_prx.bib}

\end{document}

%% file: Sections/1_Introduction.tex
A central problem in non-equilibrium quantum many-body physics is to determine whether real-time evolution admits organizing principles comparable to those that govern equilibrium statistical mechanics~\cite{Hohenberg1977}.
In equilibrium, phase transitions arise as non-analytic behavior of the free energy as external parameters are varied, thereby sharply separating distinct macroscopic regimes~\cite{cardy1984,sachdev2011quantum}.
Out of equilibrium, by contrast, there is in general no free-energy functional available to classify the dynamics, and the characterization of universal structure remains considerably less developed.

The framework of dynamical phase transitions offers one possible route toward such an organizing principle. In one common formulation, a dynamical quantum phase transition (DQPT) is defined in terms of the Loschmidt amplitude $\bra{\psi_0}e^{-i\hat{H}t}\ket{\psi_0}$. This quantity measures the overlap between the initial state $\ket{\psi_0}$ and its time-evolved counterpart after a quench with a Hamiltonian $\hat{H}$, and can be interpreted as a dynamical analogue of a partition function. Complex zeros of this partition function can give rise to non-analytic behavior in the associated return rate, in close analogy with the relation between equilibrium free energies and conventional partition functions~\cite{heyl2013}. A complementary notion classifies dynamical phases according to whether the long-time state preserves or breaks a symmetry of the post-quench Hamiltonian. In this formulation, the relevant dynamical phases are distinguished by their symmetry-breaking patterns~\cite{Sciolla2013}. Although these two notions are distinct, they convey the same basic lesson: in an appropriate thermodynamic limit, real-time dynamics can exhibit sharp transitions between qualitatively different regimes.

In this work, we use the term dynamical phase transition in this restricted sense. We do not study non-analyticities of the Loschmidt return rate. Instead, we study non-analyticities of mutual information in the large-central-charge limit of $1+1$-dimensional conformal field theories with a holographic dual. The relevant limiting procedure is therefore the semiclassical $c\to\infty$ limit, in which conformal blocks exponentiate. A dynamical phase transition occurs when, as a function of real time, the dominant conformal block — equivalently, the minimal holographic geodesic configuration — changes. The resulting non-analyticity in mutual information is therefore not simply a cusp in an arbitrary time-dependent observable, but marks a genuine reorganization of the system's entanglement structure.

This perspective naturally extends the DQPT idea to a non-local probe of correlations. Mutual information measures correlations shared between spatial regions and is insensitive to ultraviolet divergences in individual entanglement entropies. After a local quench, its time dependence is governed by the motion of interval endpoints under the conformal map to the upper half-plane. The corresponding cross-ratios become time-dependent, and critical times occur when they cross the boundaries between distinct geodesic, or equivalently conformal-block, phases. In this sense, time acts as a control parameter for a sharply defined phase diagram of mutual information.

From this perspective, our observation that the mutual information (MI) in $1+1$-dimensional conformal field theory (CFT) at large central charge undergoes multiple dynamical phase transitions is particularly compelling.
Mutual information is a non-local observable that probes the pattern of correlations shared between spatial sub-regions, and is therefore intrinsically sensitive to the redistribution of entanglement generated by a quench.
We show that this quantity undergoes multiple dynamical phase transitions in a variety of local quench protocols and provide a detailed map of the rich dynamical phase structure: there are 6 phases of the mutual information, each governed by the dominance of a given conformal block (or equivalently by a given holographic geodesic pattern).
These transitions arise in the large central charge limit $c\rightarrow\infty$, which roughly corresponds to a limit of infinitely many degrees of freedom. In this respect, the situation closely parallels conventional phase transitions, which emerge only in the thermodynamic limit.

The dynamical phases are able to explain more details of the time evolution than the widely used quasiparticle picture of post-quench entanglement dynamics.
Although the quasiparticle picture offers valuable intuition for the causal propagation of correlations, we show that the phases allow us to explain non-analytic behavior not predicted by light-like propagation from quench points.

A further noteworthy aspect of the new paradigm is the identification of the breaking and restoration of a $D_4$ symmetry which controls the presence/absence of mutual information\footnote{We would like to emphasize that this is not a dynamical understanding of a broken equilibrium symmetry~\cite{heyl2014}, but rather a genuinely dynamical symmetry that does not exist in equilibrium.}.
This suggests that distinct dynamical regimes may be characterized by an underlying symmetry principle.
Such an interpretation is highly suggestive. In equilibrium, symmetry and its spontaneous breaking furnish the foundation of the Landau paradigm, providing a unifying language for phase structure and universality.
The fact that an analogous symmetry-based characterization appears in the present non-equilibrium setting raises the possibility that entanglement dynamics in conformal quantum matter may admit a similarly organized description.
In particular, the onset or disappearance of mutual information may be viewed as the dynamical analogue of entering or leaving an ordered phase, with the associated non-analyticity marking a genuine phase boundary in time.

Our results bring together several strands of the literature that are usually discussed separately.
The first is the theory of dynamical quantum phase transitions. Since the original proposal that non-analytic behavior can occur in real-time evolution~\cite{heyl2013}, this subject has developed into a broad program aimed at identifying robust notions of phase structure away from equilibrium~\cite{Karrasch2013,Vajna2014,pozsgay2013,Andraschko2014,Halimeh2020,Halimeh2021,Osborne2024,Zunkovic2016,Homrighausen2017,Halimeh2017,Zauner-Stauber2017,Defenu2019,Uhrich2020,Halimeh2021,Corps2022,Corps2023a,Corps2023b,Mitra2025,Schmitt2015,Bhattacharya2017,Srivastav2019,DeNicola2019,Hashizume2020,Hashizume2022,Vajna2015,Schmitt2015,Huang2016,Hagymasi2019,Srivastav2019,Porta2020,Okugawa2021,Cao2025,Abeling2016,bhattacharya2017b,Lang2018a,Lang2018b,Mera2018,Corps2024,Cao2025,Zhou2018,Wang2019,Zhou2021,Hamazaki2021,Mondal2022,Kawabata2023,Mondal2023,Mondal2024,Fu2025,zhang2025,parez2026,mondkar2026,gu2026,Kosior2018a,Kosior2018b,mondkar2026,Halimeh2019,Trapin2021,Zache2019,Huang2019,Pedersen2021,Jensen2022,Halimeh2022,VanDamme2022,Mueller2023,Pomarico2023,VanDamme2023,osborne2025c,osborne2026unifiedresonantmanifoldframeworkdynamical}.
In analogy with equilibrium criticality, dynamical phase transitions can be understood through the Lee--Yang and Fisher picture of zeros and singularities~\cite{lee1952,fischer1965}, where these singularities occur on Lorentzian world sheets rather than the thermal circle.
This topic has also become experimentally accessible with observations in systems of trapped ions, cold atoms, and related platforms~\cite{jurcevic2017,nie2020,flaschner2018}.
What our work adds to this discussion is a concrete entanglement-based realization in $1+1$-dimensional large-$c$ CFT: the mutual information does not merely vary in time, but passes through a sequence of non-analytic transitions, so that the redistribution of correlations is naturally organized into dynamical phases.
Our work provides a universal starting point where finite central charge and non-conformal corrections can be calculated in specific theories of interest.

A second point of contact is the extensive literature on quantum quenches and entanglement dynamics in low-dimensional many-body systems.
Quantum quenches are a paradigmatic setting for non-equilibrium dynamics in cold-atom experiments~\cite{kinoshita2006a,greiner2002,gring2012}, and their theoretical description in conformal field theory was developed in the foundational work of Calabrese and Cardy and many later extensions~\cite{calabrese2004a,calabrese2005c,calabrese2007a,calabrese2007b,calabrese2009a,calabrese2016a}.
These developments established the basic picture that entanglement propagation can be helpfully understood with pairs of entangled quasiparticles that are emitted at the quench point, where entanglement saturation occurs when the quasiparticles have moved beyond the subsystem endpoints.
They also motivated a much broader study of integrable and near-integrable out-of-equilibrium dynamics~\cite{cardy2007,delfino2004,delfino2017,fioretto2010,dutta2015,delfino2022,polkovnikov2011,delfino2024}.

A third thread enters through holography, and in particular through AdS/BCFT.
The general AdS$_3$/CFT$_2$ correspondence and the holographic description of entanglement entropy has a long history~\cite{brown1986a,krasnov2000b,zograf1988,ryu2006b,ryu2006c,hubeny2007a}.
The extension of this framework to systems with boundaries~\cite{takayanagi2011d,fujita2011} is especially important for us because splitting and joining quenches are most naturally formulated in a BCFT language.
In this setting the boundary is represented by an end-of-the-world brane~\cite{randall1999}, and quench dynamics can be reinterpreted geometrically in terms of the dynamics of geodesics~\cite{shimaji2019b,caputa2019a,lap2024c,lap2025a,caputa2025a,Balasubramanian_2011}.
While this geometric framework is useful for visualizing the physics, it is by now well-understood that the geometry based formulas can be derived directly from two-dimensional CFTs~\cite{hartman2013a,hartman2014,asplund2015}.
Correlation functions involving operators whose dimensions scale with the central charge are known to simplify dramatically at large $c$, and in this regime Virasoro blocks exponentiate and admit a semiclassical saddle-point interpretation~\cite{zamolodchikov1984,perlmutter2014,perlmutter2015,hijano2015,fitzpatrick2016,fitzpatrick2017,fitzpatrick2017a,besken2020}.
These developments undergird the modern field-theoretic understanding of heavy operator insertions more generally~\cite{abajian2023,poland2026b,chiang2026}.
So long as the low-lying operator spectrum is sufficiently sparse~\cite{hartman2013a,hartman2014}, at large $c$ one recovers the holographic result.
Therefore we view holography as a useful language: it gives an efficient geometric way of visualizing which OPE channel dominates in a given regime and which symmetries are at play.
This well-trodden duality has also been used to probe quantum gravity where it has given insight into black hole dynamics and interiors~\cite{sully2021a,geng2022,geng2025,bao2025}.

Taken together, these strands allow a unified interpretation of non-equilibrium entanglement dynamics.
The quench literature tells us which observables and protocols are natural; the heavy-operator and semiclassical block literature tells us our dynamics are universal in the large $c$ limit; AdS/BCFT provides an efficient geometric picture of which OPE channels dominate and how boundary conditions affect phase behavior.
For the dynamical phase transitions in mutual information we consider here, these pictures fit together into a universal description of real-time criticality where dynamical phase transitions are governed by conformal-block dominance and the dynamical emergence of mutual information is governed by the breaking and restoration of a $D_4$ symmetry.
This opens the possibility of a Landau-like description of non-equilibrium physics, formulated in the familiar universal language of symmetries and order parameters.

The layout of the paper is as follows.
In Sec.~\ref{sec:section2}, we provide a lightning review of the ingredients used in our computations and references to related literature. In Sec.~\ref{sec:section3}, we focus on static mutual information on the upper half-plane (UHP), tabulating all possible phases of mutual information, what governs the phase transitions, and the $D_4$ symmetry associated with the existence of mutual information.
In Sec.~\ref{sec:section4}, we use various conformal maps to relate the dynamics of mutual information in quench scenarios to the phases on the upper half plane.
We show that mutual information undergoes multiple dynamical phase transitions explained by the quantities we construct in Sec.~\ref{sec:section3}.
In Sec.~\ref{sec:section5}, we provide a numerical exploration of finite $c$ effects. We calculate the mutual information in critical spin chains and show that the sharp transitions between distinct mutual‑informations are smoothed out as expected, while the transitions between phases with and without mutual information remain non‑analytic. In Sec.~\ref{sec:section6}, we conclude our discussion and mention some future directions.

%% file: Sections/2_Review.tex
\subsection{Two-dimensional CFT}
In one spatial dimension, many quantum critical systems realized in spin chains and cold-atom platforms flow in the infrared (IR) to $1+1$-dimensional conformal field theories.
Physically, this occurs when the correlation length becomes much larger than the lattice spacing and the long-distance theory becomes invariant under changes of scale; in two spacetime dimensions, this symmetry is generically enhanced to local conformal symmetry, which strongly constrains the universal low-energy behavior.
This framework successfully describes, for example, the critical Heisenberg chain, Luttinger liquids, and a wide class of bosonic and fermionic cold-atom systems tuned to quantum criticality~\cite{cardy1984,difrancesco1997,belavin1984,ribault2018}.
A conformal field theory is therefore a quantum field theory invariant under conformal transformations—coordinate transformations that preserve local angles.
In two space-time dimensions (2D) the algebra of such transformations is infinite-dimensional and is known as the Virasoro algebra~\cite{virasoro1970}\footnote{Many of these statements hold in higher dimensions.
For an introduction consider the following resources~\cite{simmons-duffin2016,rychkov2017}}.
Its generators $L_n$ satisfy
\be[L_m,L_n]=(m-n)L_{m+n}+\frac{c}{12}(m^3-m)\delta_{m+n,0},\ee
with an analogous anti-holomorphic copy \(\bar L_n\); the parameter $c$ is the central charge, which measures the number of degrees of freedom in the CFT.
The local operators of the theory are then organized into representations of this algebra.
A Virasoro primary operator $\phi(z,\bar z)$ is
\begin{enumerate}
    \item Annihilated by all positive modes, 
    \be L_n\phi=\bar L_n\phi=0 \quad {\rm for}~ n>0,\ee
    \item An eigenoperator of $L_0,\bar{L}_0$
    \be L_0\phi=h\phi,\,\bar{L}_0\phi=\bar{h}\phi.\ee
\end{enumerate}
The numbers $(h,\bar{h})$ are the holomorphic and anti-holomorphic conformal weights respectively.
The total scaling dimension of an operator is their sum $\Delta=h+\bar{h}$.

These operators are called primary because they are highest weight states, and all other operators – descendants – can be generated by acting on the primary with $L_{-n}$ and $\bar{L}_{-n}$.\\

2D CFTs are invariant under a subgroup of the local conformal group – generated by the Virasoro algebra – known as the global conformal group –$PSL(2,\mathbb{C})$.
This is the group of all rotations, scalings, and special conformal transformations and is generated by $L_{-1},L_0$ and $L_1$.
They act on the coordinates with M\"obius transformations
\be 
\left\{z\rightarrow \left.\frac{a  z+b}{cz+d}\, \right| a,b,c,d\in\mathbb{C}~{\rm with}~ad-bc\neq0 \right\}.
\ee
This group has 3 complex degrees of freedom and can be used to fix any three positions of a correlation function. The invariants of this group are known as cross-ratios and have the form
\be \CR(z_1,z_2,z_3,z_4)=\frac{(z_1-z_2)(z_3-z_4)}{(z_1-z_3)(z_2-z_4)}.\ee
When considering correlation functions of local operators this heavily constrains the form they can take.
1-point, 2-point, and 3-point functions are fully fixed by symmetry, the first theory-specific correlation function is the 4-point function. Consider a 4-point function of identical Virasoro primaries
\be \braket{\phi(z_1)\phi(z_2)\phi(z_3)\phi(z_4)}=\frac{g(z,\bar{z})}{|z_{12}|^{2\Delta}|z_{34}|^{2\Delta}}.\ee
Where \(z_{ij}=z_i-z_j\), \(z=\CR(z_1,z_2,z_3,z_4)\), \(\bar{z}=\CR(\bar{z}_1,\bar{z}_2,\bar{z}_3,\bar{z}_4)\), and \(g(z,\bar{z})\) is an unspecified theory dependent function.
We can figure out what the function should be in a given theory by doing an operator product expansion (OPE) which reduces the 4-point function to a sum over 3pt functions (which are fixed by conformal invariance).
Unlike in standard QFT, in CFT the OPE is not just an asymptotic expansion but convergent in a given ``OPE channel''.
For example in the $s$-channel (where $z\rightarrow0$), which converges for $|z|<1$: 
\be g(z,\bar{z})
=
\sum_p |C_{\phi\phi p}|^{\,2}\,
\mathcal F_p(z)\,\overline{\mathcal F}_p(\bar{z}),\ee

Where the sum is over primary operators $p$, $C_{\phi\phi p}$ is the OPE coefficient of the 3pt function of two insertions of $\phi$ and one of the primary $p$, and $\mathcal{F}_p(z)$ is a holomorphic function known as a conformal block\cite{difrancesco1997}.\\

Similarly in the $t$-channel (where $z \rightarrow 1$) one finds:
\be g(z,\bar{z})
=|z|^{2\Delta}\sum_p |C_{\phi\phi p}|^{\,2}\,
\mathcal F_p(1-z)\,\overline{\mathcal F}_p(1-\bar{z})\ee
Which converges for $|1-z|<1$

Crossing symmetry is the statement that these distinct channel expansions represent the same correlator, thereby imposing non-trivial consistency conditions on the operator spectrum and OPE coefficients.

\begin{figure}
    \centering
    \includegraphics[width=0.8\linewidth]{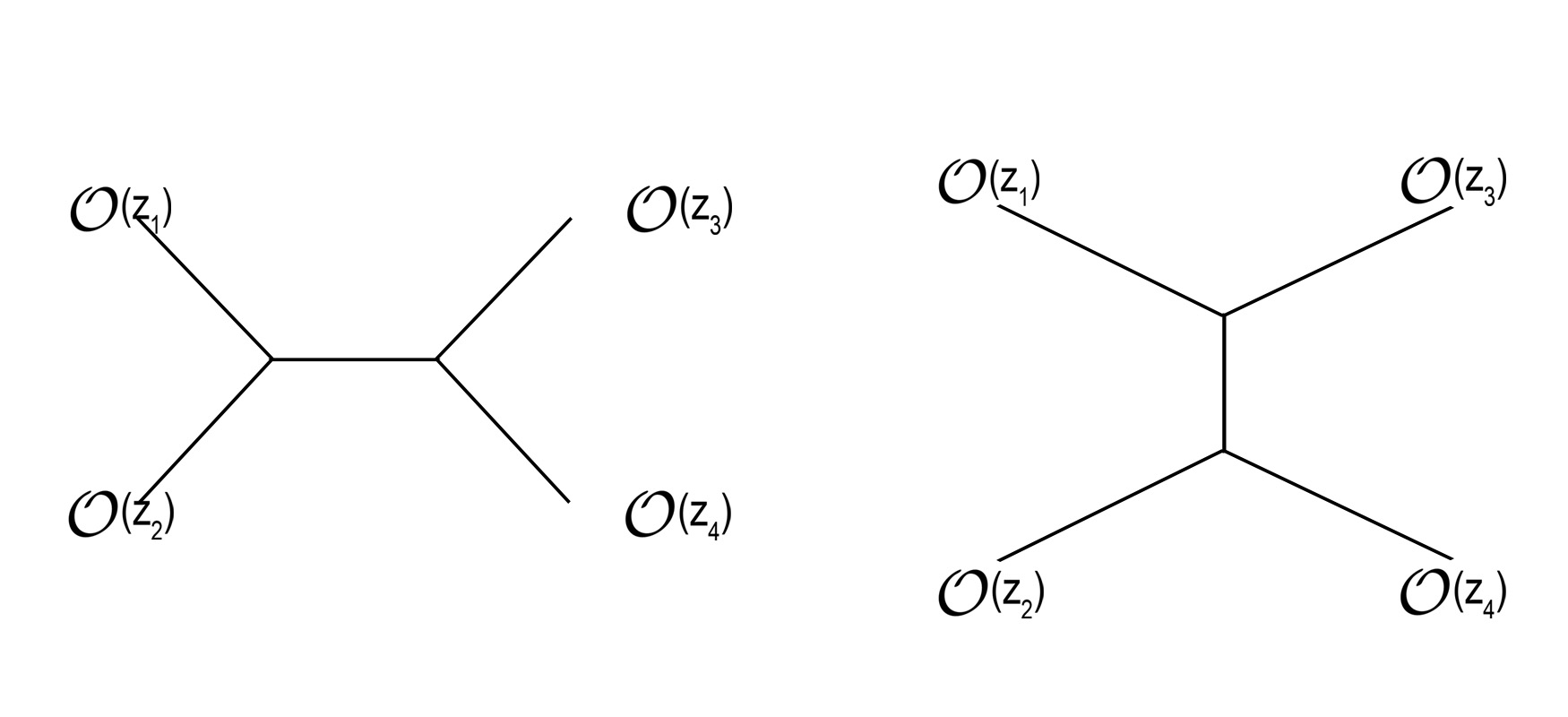}
    \caption{The expansion of a 4-point function in the $s$- and $t$-channels.}
\end{figure}

\subsection{BCFT}
A boundary conformal field theory (BCFT) is a conformal field theory on a manifold with boundaries that are conformally invariant\footnote{All of the following can be found in more detail in Cardy's Review\cite{cardy2004}}.
In 2D, the condition that no energy/momentum flows over the boundary enforces that the holomorphic and anti-holomorphic copy of the stress-tensor are equivalent at the boundary.
This breaks the two chiral copies of the Virasoro algebra, holomorphic and anti-holomorphic, to one diagonal Virasoro algebra.
For diagonal CFTs a consequence of this is that there is a direct correspondence between conformally invariant boundary conditions – Cardy states – and the primary operators of the CFT. We treat all possible boundary conditions by parameterizing them by the boundary entropy they induce.
The inner product of a conformal boundary condition $B$ and the vacuum gives the Ludwig-Affleck~\cite{affleck1991} g-function \(\ g=\braket{0|B}\), which is a real number that measures the number of degrees of freedom of a given boundary condition.
As such it contributes to the entropy of the BCFT adding a term
\begin{equation}
    S_{bdy}=\ln(g).
\end{equation}

Rather than calculating the allowed boundary conditions of a specific theory and enumerating them, we provide all equations in terms of $g$ which can be calculated for a specific choice of conformally invariant boundary condition\footnote{The question of what conformally invariant boundary conditions a theory allows is a subtle one. The Cardy procedure only works for diagonal CFTs which is a small subset of all CFTs. In addition certain conformally invariant boundary conditions insert operators into the path-integral and thus do not compute entanglement entropy but some closely related quantity~\cite{agia2022}. }.
As such we frequently use a parameterization that arises more naturally in holography:
\be 
S_{bdy}=\ln(g)=\frac{c}{6}\arctanh(T)=\frac{c}{6}\ln(k),
\label{eq:S-bdy}
\ee
Where $T$ is the energy on the brane\footnote{Strictly $S_{bdy}=\frac{c}{6}\arctanh(R_{AdS}T)$ but we set the $AdS$ radius to 1 for convenience.} corresponding to the boundary condition (see \ref{sec:holographyB}), and the $k$ parameterization was introduced in~\cite{shimaji2019b}.

\subsection{Holography}
In two-dimensional CFT, entanglement entropy is commonly computed using the path-integral formalism together with the replica trick~\cite{calabrese2016a}.
The von Neumann entropy of a subsystem \(A\) is
\be
S_A=-\Tr(\rho_A\ln\rho_A)=\lim_{n\rightarrow 1}\partial_n \Tr(\rho_A^n),
\ee
where the \(n\)-th moment of the reduced density matrix, \(\Tr(\rho_A^n)\), is represented by a path integral on an \(n\)-sheeted Riemann surface cyclically branched over \(A\).
Instead of looking at a complicated Riemann surface, we can insert twist operators $\sigma_n(x)$ at the endpoints $x_1$ and $x_2$ of \(A\) in the complex plane~\cite{cardy2007,calabrese2005c} to encode the branch-point structure.
For the vacuum state of a CFT on the infinite line, the result for a single interval \(A=[0,l]\), is the classic result~\cite{holzhey1994}
\be
\label{eq:vacuum EE}
S_A=\frac{c}{3}\ln\!\left(\frac{l}{\epsilon_\mathrm{UV}}\right),
\ee
where \(c\) is the central charge and \(\epsilon_\mathrm{UV}\) is an ultra-violet cutoff.

In a holographic CFT, this expression admits a geometric interpretation: It is precisely the length of the corresponding geodesic in hyperbolic space.
Ryu and Takayanagi therefore conjectured that the entanglement entropy of a boundary region is given by the length of the bulk geodesic\footnote{For vacuum states there is typically a unique such geodesic, whereas for thermal states or theories with boundaries several candidate geodesics may exist. The physically relevant one is the shortest.} homologous (continuously-deformable) to that region~\cite{ryu2006c}, viz. the shortest path connecting the endpoints of the subsystem through the hyperbolic bulk space.

This leads to predictions of sharp phase behavior.
When calculating the entanglement entropy of two disjoint intervals $X=[z_1,z_2],Y=[z_3,z_4]$, there is a phase where the shortest geodesics connect $[z_1,z_2]$ and $[z_3,z_4]$ and there is a ``rainbow''-like phase where they connect $[z_1,z_4]$ and $[z_2,z_3]$.
See Fig.~\ref{fig:vacuum MI geodesics}.
Such phase behavior arises because CFTs with holographic duals are a special subset of CFTs~\cite{hartman2013a}:
where the central charge $c$ is large, and the number of Virasoro primaries with low scaling dimension $\Delta$ is much smaller than the number of those with large scaling dimension~\cite{cardy1986}.
In such theories, different conformal blocks dominate in different OPE channels leading to universal, non-trivial phase behavior.
From this perspective, the Ryu-Takayanagi prescription can be viewed as an elegant geometric representation of otherwise intricate OPE decompositions\footnote{Unlike a four-point function, an eight-point function admits four inequivalent topologies~\cite{fortin2020}, which are the higher-point analogues of the two topologically inequivalent decompositions of the six-point function: the comb\cite{rosenhaus2019} and snowflake channels.}.

Most CFTs realized in spin chains or cold-atom experiments, however, have relatively small central charge.
In that case, \(1/c\) corrections become important, additional conformal blocks contribute, and the sharp holographic transitions are smoothed out.
We return to this point in Sec.~\ref{sec:section5}.

\subsection{Holography with Boundaries}
\label{sec:holographyB}
The central question of AdS/BCFT is how one extends holography to a conformal field theory defined on a space with a boundary~\cite{takayanagi2011d,fujita2011}.
The basic idea is to truncate the bulk spacetime by introducing an end-of-the-world brane \(Q\), together with the appropriate Gibbons-Hawking-York boundary term in the Euclidean gravitational action
\be I_E=\frac{-1}{16\pi G_N}\int_{N} \sqrt{g}(R-2\Lambda)-\frac{1}{8\pi G_N}\int_{Q} \sqrt{h}(K-T).\ee
Here \(h\) is the induced metric on \(Q\), \(K\) its extrinsic curvature, and \(T\) the brane tension.
Different conformally invariant boundary conditions of the BCFT are encoded by different values of \(T\).
The corresponding bulk description imposes Neumann boundary conditions on the end-of-the-world brane, so that its extrinsic curvature is fixed by the tension, \(K \propto T\), as in the standard braneworld picture~\cite{randall1999a}.

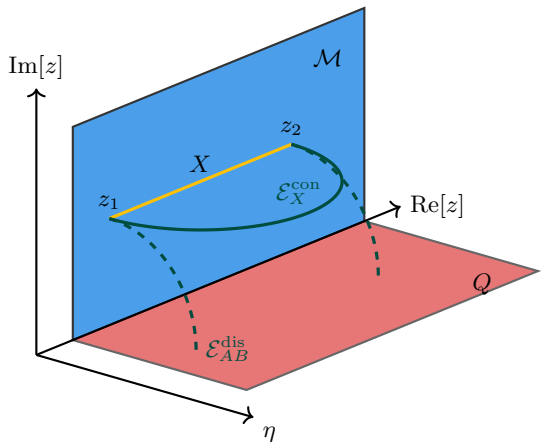
\begin{figure}
    \centering
    \tdplotsetmaincoords{70}{-50}
    \begin{tikzpicture}[
            tdplot_main_coords,
            scale=1.5,
            axis/.style={->,black,thick},
            line/.style={very thick, color=black},
            upper/.style={thick, fill=color_upper, opacity=0.8},
            lower/.style={thick, fill=color_lower, opacity=0.8},
            boundary/.style={thick, fill=color_boundary, opacity=0.6},
            sub/.style={very thick, color=color_sub, opacity=1},
            geo/.style={very thick, color=color_geo, opacity=1}]
            
        \draw[axis] (-0.5,4,0) -- (-0.5,4,2.5) node[above]{$\mathrm{Im}{[z]}$};
        \draw[axis] (-0.5,4,0) -- (4.5,4,0) node[right]{$\mathrm{Re}{[z]}$};
        \draw[upper] (0,4,0) -- (4,4,0) -- (4,4,2) -- (0,4,2) -- cycle;
       
        \draw[boundary] (0,2,0) -- (4,2,0) -- (4,4,0) -- (0,4,0) -- cycle;
       
        \draw[sub] (0.85,0,1.03) node[black, above]{$Q$};
        \draw[sub] (3.5,4,1.5) node[black, above]{$\mathcal{M}$};
        \draw[sub] (0.5,4,1) node[black, above]{$z_1$} -- (1.75,4,1) node[black, above]{$X$} -- (3,4,1) node[black, above]{$z_2$};
        \draw[geo, variable=\t, domain=0:-pi, samples=200]
        plot({1.75 + 1.25*cos(\t r)}, {4 + 1.25*sin(\t r)}, {1});
        \draw[sub] (0.5,3,0.23) node[geo, below right]{$\mathcal{E}_{AB}^{\rm dis}$};
        \draw[sub] (0.6,1.9,1.7) node[geo]{$\mathcal{E}_X^{\rm con}$};
        \draw[dashed,geo, variable=\t, domain=0:-1*pi/2, samples=200] plot({0.5},{4+sin(\t r)},{cos(\t r)});
        \draw[dashed,geo, variable=\t, domain=0:-1*pi/2, samples=200] plot({3},{4+sin(\t r)},{cos(\t r)});
        \draw[axis] (-0.5,4,0) -- (-0.5,1.5,0) node[below right]{$\eta$};
    \end{tikzpicture}
    \caption{A sketch of the extremal geodesics in the presence of a boundary $Q$ (red). In addition to the connected geodesic $\mathcal{E}_X^{\rm con}$ (solid black), one also has to take into account the disconnected geodesic $\mathcal{E}_X^{\rm dis}$ (dashed black) that ends on the boundary $Q$, viz. the end-of-the-world brane.}
    \label{fig:vacuum_RT_formula}
\end{figure}

The addition of an end-of-the-world brane means geodesics now come in two classes (see Fig.~\ref{fig:vacuum_RT_formula}): connected (attached to the interval endpoints) or disconnected (going from each interval endpoint to the brane). The result is that a given boundary condition - expressed via the brane tensions $T$ - will shift phase behavior by a constant amount in time.

\subsection{Mutual Information}
As is evident from eq.~\eqref{eq:vacuum EE} the entanglement entropy depends on the UV-cutoff $\epsilon_\mathrm{UV}$ and will diverge in the $\epsilon_\mathrm{UV} \rightarrow 0$ limit.
To get rid of this divergence we can consider a difference of entanglement entropies, where for splitting quenches one often considers $\Delta S_A(t) = S_A(t) - S_0$ (where $S_0$ is the entanglement entropy of the unquenched system).
Another option to remove this divergence is to consider the so-called mutual information
\begin{equation}
    I_{A:B} = S_A + S_B - S_{A \cup B},
\end{equation}
This quantity measures the amount of information shared between the regions $A$ and $B$, excluding correlations with the complement of $A \cup B$. In this sense, it captures the information uniquely shared by $A$ and $B$ and provides a measure of bipartite entanglement. Higher-party generalizations of this notion have recently been explored ~\cite{balasubramanian2024signalsmultipartyentanglementholography,balasubramanian2026timeevolutionmultipartyentanglement,Balasubramanian_2026}. Mutual information will be the main focus of this paper. We analyze its dynamical behavior following a local quench in terms of phases associated with the disjoint subsystem $A \cup B$. A detailed introduction to these phases is given in Sec.~\ref{sec:section3}.

\subsection{Worldsheets for Various Quenches}
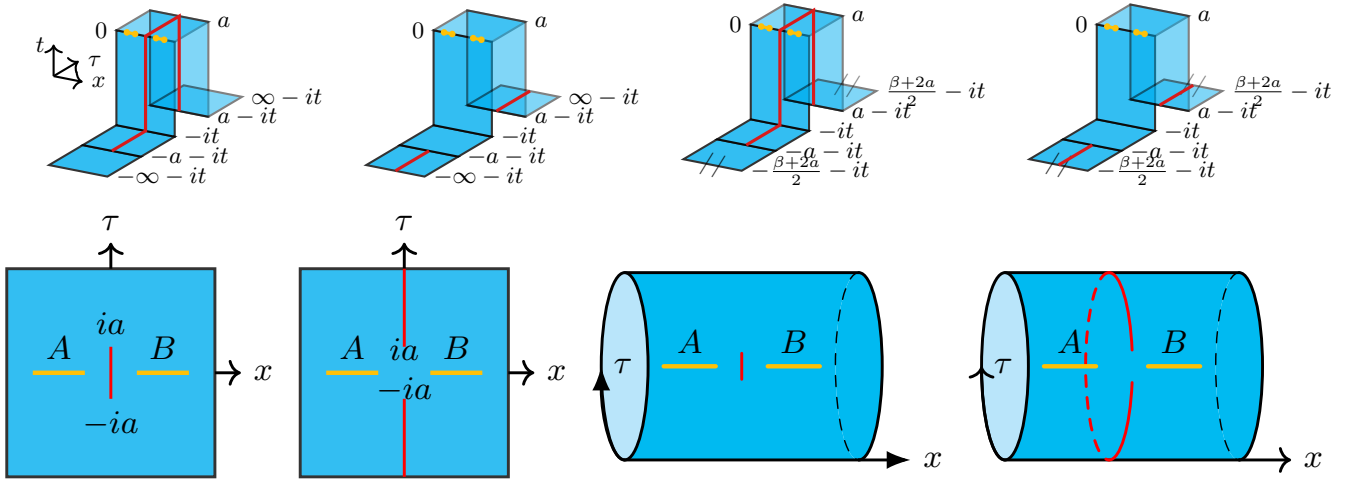
\begin{figure*}
    \centering
    \resizebox{\linewidth}{!}{\input{Plots/Worldsheets.tex}}
     \resizebox{\linewidth}{!}{\input{Plots/FlatWorldsheets.tex}}
    \caption{First Row: The world-sheets for a) splitting quench, b) joining quench c) splitting quench at finite temperature d) joining quench at finite temperature.\\
    Second Row: The corresponding world-sheets at $t=0$.}
    \label{fig:Worldsheets}
\end{figure*}

Consider the quantum state $\psi(t)$ induced by a quantum quench: where the vacuum of Hamiltonian $H_1$ is time evolved with Hamiltonian $H_2$.\\
We can prepare the vacuum state of $H_1$
\be \ket{0_1}=\lim_{\tau\rightarrow\infty}e^{-iH_1(-it)}\ket{\phi},\ee
by evolving an arbitrary state for infinite euclidean time $\tau=-it$.
The post-quench state $\psi(t)$ then is
\be \psi(t)=e^{-i H_2t}\ket{0_1}.\ee
For splitting and joining quenches, where only the spatial support of the Hamiltonian is changed, the state develops ultraviolet divergences unless a short-distance regulator, $a$, is introduced~\cite{calabrese2007b}.
We therefore work with the regulated state
\be \psi_a(t)=e^{-i H_2t}e^{-H_2a}\ket{0_1}.\ee
Consider the expectation value of twist operators placed at $x_1$ and $x_2$
\be \Tr(\rho_A^n(t))=\braket{\psi(t)|\sigma_n(x_1)\sigma_{\bar{n}}(x_2)|\psi(t)}.\ee
The path integral representation (the world-sheet) of this density matrix is shown in the first two columns of Fig.~\ref{fig:Worldsheets} for the splitting and joining quench, respectively.

A thermal initial state can be treated similarly using the thermofield-double, or thermal-circle, formalism~\cite{takahashi1996,matsubara1955,maldacena2003a,hartman2015,cottrell2019a}.
One purifies the thermal density matrix by introducing a second copy of the Hilbert space, displaced by \(\beta/2\) in Euclidean time, and then traces over the auxiliary copy.
This construction produces a cylinder of circumference \(\beta\) with open bra and ket boundaries that prepare the thermal state before the quench.
The corresponding world-sheets are shown in the right two columns of Fig.~\ref{fig:Worldsheets}.

In each of these quench backgrounds, we follow the standard procedure~\cite{calabrese2016a,shimaji2019b,caputa2019a}.
We first compute \(\Tr(\rho_A^n)\) in Euclidean time on the relevant world-sheet, then map the geometry conformally to the upper half-plane, where the calculation simplifies, and finally analytically continue back to real time.
Every world-sheet we consider is conformally equivalent to the upper half-plane. In this sense, the upper half-plane ``uniformizes'' the various world-sheets: Euclidean correlators of twist operators on the upper half-plane encode the Lorentzian dynamics of entanglement in these distinct quench protocols.

\subsection{Conformal Mapping}
Since we are working with a 1+1 dimensional CFT we can apply conformal transformations to the CFT, which will not change the entanglement entropy.
For a world-sheet with complex coordinate $w = x + i\tau$ this transformation is denoted by
\begin{equation}
\label{eq:conformal map}
    z = f(w),
\end{equation}
where $z$ is the complex coordinate on the target domain and $f$ is the conformal map.

As shown in~\cite{roberts2012a} for holographic CFTs this implies a transformation of the AdS coordinates $(x,\tau,\xi) \sim (w,\xi)$ (where $\xi$ is the bulk coordinate of the AdS space dual to the ``physical'' CFT) to some target AdS space with coordinates $(\mathrm{Im}(z),\mathrm{Re}(z),\eta) \sim (z,\eta)$, which is given by
\begin{equation}
    \begin{aligned}
        z &= f(w) - \frac{2 \xi^2 (f')^2 (\bar{f}'')}{4 |f'|^2 + \xi^2 |f''|^2}\\
        \bar{z} &= \bar{f}(\bar{w}) - \frac{2 \xi^2 (\bar{f}')^2 (f'')}{4 |f'|^2 + \xi^2 |f''|^2}\\
        \eta &= \frac{4 \xi (f' \bar{f}')^{3/2}}{4 |f'|^2 + \xi^2 |f''|^2},
    \end{aligned}
\end{equation}
where primes denote derivatives with respect to the complex coordinate $w$.

%% file: Plots/Worldsheets.tex
\centering
\begin{tabular}{cccc}
        \centering
        \tdplotsetmaincoords{70}{30}
        \begin{tikzpicture}[
            tdplot_main_coords,
            scale=0.45,
            axis/.style={->,black,thick},
            ket/.style={thick, fill=cyan, opacity=0.8},
            bra/.style={thick, fill=cyan, opacity=0.5},
            line/.style={very thick, color=color_boundary},
            sub/.style={very thick, color=color_sub, opacity=1}]
            
            \draw[axis] (-1,2,2) -- (0,2,2) node[anchor=west]{$x$};
            \draw[axis] (-1,2,2) -- (-1,3.5,2) node[anchor=west]{$\tau$};
            \draw[axis] (-1,2,2) -- (-1,2,3) node[anchor=east]{$t$};
        
            \draw[ket] (0,0,0) -- (2,0,0) node[right,opacity=1]{$-\infty-it$} -- (2,2,0) node[right,opacity=1]{$-a-it$} -- (0,2,0) node[above left,opacity=1]{} -- cycle;
            \draw[ket] (0,2,0) -- (2,2,0) -- (2,4,0) node[right,opacity=1]{$-it$} -- (0,4,0) -- cycle;
            \draw[ket] (0,4,0) -- (2,4,0) -- (2,4,3) -- (0,4,3) node[left,opacity=1]{$0$} -- cycle;
        
            \draw[bra] (0,4,3) -- (2,4,3) -- (2,6,3) node[right, opacity=1]{$a$} -- (0,6,3) -- cycle;
            \draw[bra] (0,6,3) -- (2,6,3) -- (2,6,0) -- (0,6,0) -- cycle;
            \draw[bra] (0,6,0) -- (2,6,0) -- (2,6,0) -- (0,6,0) -- cycle;
            \draw[bra] (0,6,0) -- (2,6,0) node[right,opacity=1]{$a-it$} --(2,8,0) node[right,opacity=1]{$\infty-it$} -- (0,8,0) -- cycle;
        
            \draw[line] (1,2,0) -- (1,4,0) -- (1,4,3) -- (1,6,3) -- (1,6,0);

            \draw[sub, arrows = {Circle[scale=0.4]-Circle[scale=0.4]}] (0.25,4,3) -- (0.75,4,3);
            \draw[sub, arrows = {Circle[scale=0.4]-Circle[scale=0.4]}] (1.25,4,3) -- (1.75,4,3);
        \end{tikzpicture}
&

        \centering
        \tdplotsetmaincoords{70}{30}
        \begin{tikzpicture}[
            tdplot_main_coords,
            scale=0.45,
            axis/.style={->,black,thick},
            ket/.style={thick, fill=cyan, opacity=0.8},
            bra/.style={thick, fill=cyan, opacity=0.5},
            line/.style={very thick, color=color_boundary},
            sub/.style={very thick, color=color_sub, opacity=1}]
            

            \draw[ket] (0,0,0) -- (2,0,0) node[right,opacity=1]{$-\infty-it$} -- (2,2,0) node[right,opacity=1]{$-a-it$} -- (0,2,0) node[above left,opacity=1]{} -- cycle;
            \draw[ket] (0,2,0) -- (2,2,0) -- (2,4,0) node[right,opacity=1]{$-it$} -- (0,4,0) -- cycle;
            \draw[ket] (0,4,0) -- (2,4,0) -- (2,4,3) -- (0,4,3) node[left,opacity=1]{$0$} -- cycle;
        
            \draw[bra] (0,4,3) -- (2,4,3) -- (2,6,3) node[right, opacity=1]{$a$} -- (0,6,3) -- cycle;
            \draw[bra] (0,6,3) -- (2,6,3) -- (2,6,0) -- (0,6,0) -- cycle;
            \draw[bra] (0,6,0) -- (2,6,0) -- (2,6,0) -- (0,6,0) -- cycle;
            \draw[bra] (0,6,0) -- (2,6,0) node[right,opacity=1]{$a-it$} --(2,8,0) node[right,opacity=1]{$\infty-it$} -- (0,8,0) -- cycle;
        
            \draw[line] (1,0,0) -- (1,2,0);
            \draw[line] (1,6,0) -- (1,8,0);

            \draw[sub, arrows = {Circle[scale=0.4]-Circle[scale=0.4]}] (0.25,4,3) -- (0.75,4,3);
            \draw[sub, arrows = {Circle[scale=0.4]-Circle[scale=0.4]}] (1.25,4,3) -- (1.75,4,3);
        \end{tikzpicture}
&
        \centering
        \tdplotsetmaincoords{70}{30}
        \begin{tikzpicture}[
            tdplot_main_coords,
            scale=0.45,
            axis/.style={->,black,thick},
            ket/.style={thick, fill=cyan, opacity=0.8},
            bra/.style={thick, fill=cyan, opacity=0.5},
            line/.style={very thick, color=color_boundary},
            sub/.style={very thick, color=color_sub, opacity=1}]
            

            \draw[ket] (0,0,0) -- node[sloped]{//} (2,0,0)  node[right,opacity=1]{$-\frac{\beta+2a}{2}-it$} -- (2,2,0) node[right,opacity=1]{$-a-it$} -- (0,2,0) node[above left,opacity=1]{} -- cycle;
            \draw[ket] (0,2,0) -- (2,2,0) -- (2,4,0) node[right,opacity=1]{$-it$} -- (0,4,0) -- cycle;
            \draw[ket] (0,4,0) -- (2,4,0) -- (2,4,3) -- (0,4,3) node[left,opacity=1]{$0$} -- cycle;
        
            \draw[bra] (0,4,3) -- (2,4,3) -- (2,6,3) node[right, opacity=1]{$a$} -- (0,6,3) -- cycle;
            \draw[bra] (0,6,3) -- (2,6,3) -- (2,6,0) -- (0,6,0) -- cycle;
            \draw[bra] (0,6,0) -- (2,6,0) -- (2,6,0) -- (0,6,0) -- cycle;
            \draw[bra] (0,6,0) -- (2,6,0) node[right,opacity=1]{$a-it$} --(2,8,0) node[right,opacity=1]{$\frac{\beta+2a}{2}-it$} 
            -- node[sloped]{//} (0,8,0) -- cycle;

            \draw[line] (1,2,0) -- (1,4,0) -- (1,4,3) -- (1,6,3) -- (1,6,0);

            \draw[sub, arrows = {Circle[scale=0.4]-Circle[scale=0.4]}] (0.25,4,3) -- (0.75,4,3);
            \draw[sub, arrows = {Circle[scale=0.4]-Circle[scale=0.4]}] (1.25,4,3) -- (1.75,4,3);
        \end{tikzpicture}
        &
        \centering
        \tdplotsetmaincoords{70}{30}
        \begin{tikzpicture}[
            tdplot_main_coords,
            scale=0.45,
            axis/.style={->,black,thick},
            ket/.style={thick, fill=cyan, opacity=0.8},
            bra/.style={thick, fill=cyan, opacity=0.5},
            line/.style={very thick, color=color_boundary},
            sub/.style={very thick, color=color_sub, opacity=1}]
            

            \draw[ket] (0,0,0) -- node[sloped]{//} (2,0,0)  node[right,opacity=1]{$-\frac{\beta+2a}{2}-it$} -- (2,2,0) node[right,opacity=1]{$-a-it$} -- (0,2,0) node[above left,opacity=1]{} -- cycle;
            \draw[ket] (0,2,0) -- (2,2,0) -- (2,4,0) node[right,opacity=1]{$-it$} -- (0,4,0) -- cycle;
            \draw[ket] (0,4,0) -- (2,4,0) -- (2,4,3) -- (0,4,3) node[left,opacity=1]{$0$} -- cycle;
        
            \draw[bra] (0,4,3) -- (2,4,3) -- (2,6,3) node[right, opacity=1]{$a$} -- (0,6,3) -- cycle;
            \draw[bra] (0,6,3) -- (2,6,3) -- (2,6,0) -- (0,6,0) -- cycle;
            \draw[bra] (0,6,0) -- (2,6,0) -- (2,6,0) -- (0,6,0) -- cycle;
            \draw[bra] (0,6,0) -- (2,6,0) node[right,opacity=1]{$a-it$} --(2,8,0) node[right,opacity=1]{$\frac{\beta+2a}{2}-it$} 
            -- node[sloped]{//} (0,8,0) -- cycle;

            \draw[line] (1,0,0) -- (1,2,0);
            \draw[line] (1,6,0) -- (1,8,0);

            \draw[sub, arrows = {Circle[scale=0.4]-Circle[scale=0.4]}] (0.25,4,3) -- (0.75,4,3);
            \draw[sub, arrows = {Circle[scale=0.4]-Circle[scale=0.4]}] (1.25,4,3) -- (1.75,4,3);
        \end{tikzpicture}        
\end{tabular}

   

%% file: Plots/FlatWorldsheets.tex
 \begin{tabular}{c c c c}

        \begin{tikzpicture}[
                axis/.style={->,black,thick},
                scale=0.5,
                upper/.style={thick, fill=cyan, opacity=0.8},
                boundary/.style={thick, fill=color_boundary, opacity=1},
                sub/.style={very thick, color=color_sub, opacity=1}]
                
            \draw[axis] (2,0) -- (2.5,0) node[right]{$x$};
            \draw[axis] (0,2) -- (0,2.5) node[above]{$\tau$};
        
            \draw[upper] (-2,-2) -- (2,-2) -- (2,2) -- (-2,2) -- cycle;
            \draw[boundary, red] (0,-0.5) node[below, black]{$-ia$} -- (0,0.5) node[above, black]{$ia$};
        
            \draw[sub] (0.5,0) -- (1,0) node[above, black]{$B$} -- (1.5,0);
            \draw[sub] (-1.5,0) -- (-1,0) node[above, black]{$A$} -- (-0.5,0);
        \end{tikzpicture}
    &
        \begin{tikzpicture}[
                axis/.style={->,black,thick},
                scale=0.5,
                upper/.style={thick, fill=cyan, opacity=0.8},
                boundary/.style={thick, fill=color_boundary, opacity=1},
                sub/.style={very thick, color=color_sub, opacity=1}]
                
            \draw[axis] (2,0) -- (2.5,0) node[right]{$x$};
            \draw[axis] (0,2) -- (0,2.5) node[above]{$\tau$};
        
            \draw[upper] (-2,-2) -- (2,-2) -- (2,2) -- (-2,2) -- cycle;
            \draw[boundary, red] (0,-2) -- (0,-0.8) node[above, black]{$-ia$} -- (0,-0.5);
            \draw[boundary, red] (0,2) -- (0,0.9) node[below, black]{$ia$} -- (0,0.5);
        
            \draw[sub] (0.5,0) -- (1,0) node[above, black]{$B$} -- (1.5,0);
            \draw[sub] (-1.5,0) -- (-1,0) node[above, black]{$A$} -- (-0.5,0);
        \end{tikzpicture}
&
\begin{tikzpicture}[axis/.style={->,black,thick},
                scale=0.5,
                upper/.style={thick, fill=cyan, opacity=0.8},
                boundary/.style={thick, color=color_boundary, opacity=1},
                sub/.style={very thick, color=color_sub, opacity=1},line cap=round,line join=round,>=Latex,rotate=90]

    \coordinate (O) at (3.3,-0.5);   
    

        \def\H{4.5}      
        \def\R{1.8}      
        \def\Ry{0.45}    
    
    \fill[cyan!80] (-\R,-\H/2) --
                   (-\R,\H/2)
                   arc[start angle=180,end angle=360,x radius=\R,y radius=\Ry]
                   -- (\R,-\H/2)
                   arc[start angle=0,end angle=180,x radius=\R,y radius=\Ry];
    
    \fill[cyan!25] (0,\H/2) ellipse[x radius=\R,y radius=\Ry];
    
    \fill[cyan!80] (0,-\H/2) ellipse[x radius=\R,y radius=\Ry];
    
    \draw[thick] (-\R,-\H/2) -- (-\R,\H/2);
    \draw[thick] (\R,-\H/2) -- (\R,\H/2);
    
    \draw[thick] (0,\H/2) ellipse[x radius=\R,y radius=\Ry];
    
    \draw[thick] (-\R,-\H/2) arc[start angle=180,end angle=360,x radius=\R,y radius=\Ry];
    \draw[dashed]  (\R,-\H/2) arc[start angle=0,end angle=180,x radius=\R,y radius=\Ry];
      
    \draw[boundary,red] (-.25,0) -- (0.25,0) node[above, black]{};

    \draw[axis] (-\R,-\H/2) -- (-\R,-\H/2-1) node[right] {$x$};
    \draw[axis] (-\R,\H/2) arc(0:90:-\R cm and \Ry cm) node[right] {$\tau$};

    \draw[sub] (0,0.5) -- (0,1) node[above, black]{$A$} -- (0,1.5);
    \draw[sub] (0,-1.5) -- (0,-1) node[above, black]{$B$} -- (0,-0.5);
    
\end{tikzpicture}
&    
\begin{tikzpicture}[axis/.style={->,black,thick},
                scale=0.5,
                upper/.style={thick, fill=cyan, opacity=0.8},
                boundary/.style={thick, color=color_boundary, opacity=1},
                sub/.style={very thick, color=color_sub, opacity=1},line cap=round,line join=round,
                rotate=90]

    \coordinate (O) at (3.3,-0.5);   
    
    
        \def\H{4.5}      
        \def\R{1.8}      
        \def\Ry{0.45}    
    
    \fill[cyan!80] (-\R,-\H/2) --
                   (-\R,\H/2)
                   arc[start angle=180,end angle=360,x radius=\R,y radius=\Ry]
                   -- (\R,-\H/2)
                   arc[start angle=0,end angle=180,x radius=\R,y radius=\Ry];
    
    \fill[cyan!25] (0,\H/2) ellipse[x radius=\R,y radius=\Ry];
    
    \fill[cyan!80] (0,-\H/2) ellipse[x radius=\R,y radius=\Ry];
    
    \draw[thick] (-\R,-\H/2) -- (-\R,\H/2);
    \draw[thick] (\R,-\H/2) -- (\R,\H/2);
    
    \draw[thick] (0,\H/2) ellipse[x radius=\R,y radius=\Ry];
    
    \draw[thick] (-\R,-\H/2) arc[start angle=180,end angle=360,x radius=\R,y radius=\Ry];
    \draw[dashed]  (\R,-\H/2) arc[start angle=0,end angle=180,x radius=\R,y radius=\Ry];
      
    \draw[boundary,red] (-\R,.25) arc[start angle=180,end angle=260,x radius=\R,y radius=\Ry];
    \draw[boundary,red] (\R,.25) arc[start angle=360,end angle=280,x radius=\R,y radius=\Ry];
    \draw[boundary,red,dashed]  (\R,.25) arc[start angle=0,end angle=180,x radius=\R,y radius=\Ry];

    \draw[axis] (-\R,-\H/2) -- (-\R,-\H/2-1) node[right] {$x$};
    \draw[axis] (-\R,\H/2) arc(0:90:-\R cm and \Ry cm) node[right] {$\tau$};

    \draw[sub] (0,0.5) -- (0,1) node[above, black]{$A$} -- (0,1.5);
    \draw[sub] (0,-1.5) -- (0,-1) node[above, black]{$B$} -- (0,-0.5);  
\end{tikzpicture}
\end{tabular}

%% file: Sections/3_Classification_of_Phases.tex
As explained in Sec.~\ref{sec:section2}, the various quench protocols we consider can all be conformally mapped to the upper half-plane (UHP). Since the physics of a CFT is conformally invariant, this means that the phase structure on the upper half-plane will govern the dynamics of the different quench protocols.
As such in this section we will focus solely on the static phase structure on the upper half-plane.
In Sec.~\ref{sec:section4} we will investigate real-time dynamics induced by pulling back these phases to the world-sheet.

When we are far away from the boundary the physics is the same as in CFT on the plane. So let's first review the two possible phases for the entanglement entropy $S_{X\cup Y}$ of two disjoint line segments $X=[z_1,z_2]$ and $Y=[z_3,z_4]$ of the plane~\cite{hartman2013a}.
The trace of the reduced density matrix on the plane for two line segments can be written as a 4-point function of twist operators
\be
\begin{aligned}
\Tr \rho_{X\cup Y}^n&=\braket{\sigma_n(z_1)\sigma_{\bar{n}}(z_2)\sigma_n(z_3)\sigma_{\bar{n}}(z_4)}\\
&=c_n \frac{g_n(x,\bar{x})}{(|z_{12}||z_{34}|)^{\frac{c}{6}(n-\frac{1}{n})}} \end{aligned} \ee
Where $g_n(x,\bar{x})$ is an $n$-dependent, theory-specific function, and the cross-ratio is $x=(z_{12}z_{34})/(z_{13}z_{24})$. 
As in Sec.~\ref{sec:section2} we take the s- ($x\rightarrow0$) and t- ($x\rightarrow 1$) channel expansions, and make the standard holographic assumptions (see Sec.~\ref{sec:section2})
\be S_{X\cup Y}=\begin{cases}
    \frac{c}{3}\ln\frac{x}{\epsilon}&\text{s-channel}\\
    \frac{c}{3}\ln\frac{1-x}{\epsilon}&\text{t-channel}
\end{cases}\ee
 
\begin{figure}
\resizebox{.9\linewidth}{!}{
\begin{tabular}{c c}
    \tdplotsetmaincoords{70}{-50}
    \begin{tikzpicture}[
            tdplot_main_coords,
            scale=0.9,
            axis/.style={->,black,thick},
            line/.style={very thick, color=black},
            upper/.style={thick, fill=color_upper, opacity=0.8},
            lower/.style={thick, fill=color_lower, opacity=0.8},
            boundary/.style={thick, fill=color_boundary, opacity=0.6},
            sub/.style={very thick, color=color_sub, opacity=1},
            geo/.style={very thick, color=color_geo, opacity=1}]
            
        \draw[axis] (-0.5,4,0) -- (-0.5,1.5,0) node[below right]{$\eta$};
        \draw[axis] (-0.5,4,0) -- (-0.5,4,3.5) node[above]{$\mathrm{Im}{[z]}$};
        \draw[axis] (-0.5,4,0) -- (4.5,4,0) node[right]{$\mathrm{Re}{[z]}$};
        \draw[upper] (0,4,0) -- (4,4,0) -- (4,4,3) -- (0,4,3) -- cycle;
        \draw[sub] (0.5,4,1) node[black, above]{$z_1$} -- (1.5,4,1) node[black, above]{$z_2$};
        \draw[sub] (2.5,4,1) node[black, above]{$z_3$} -- (3.5,4,1) node[black, above]{$z_4$};
        \draw[geo, variable=\t, domain=0:-1*pi, samples=200] plot({1+cos(\t r)/2},{4+sin(\t r)/2},{1});
        \draw[geo, variable=\t, domain=0:-1*pi, samples=200] plot({3+cos(\t r)/2},{4+sin(\t r)/2},{1});

    \end{tikzpicture}&
    \tdplotsetmaincoords{70}{-50}
    \begin{tikzpicture}[
            tdplot_main_coords,
            scale=0.9,
            axis/.style={->,black,thick},
            line/.style={very thick, color=black},
            upper/.style={thick, fill=color_upper, opacity=0.8},
            lower/.style={thick, fill=color_lower, opacity=0.8},
            boundary/.style={thick, fill=color_boundary, opacity=0.6},
            sub/.style={very thick, color=color_sub, opacity=1},
            geo/.style={very thick, color=color_geo, opacity=1}]
            
        \draw[axis] (-0.5,4,0) -- (-0.5,1.5,0) node[below right]{$\eta$};
        \draw[axis] (-0.5,4,0) -- (-0.5,4,3.5) node[above]{$\mathrm{Im}{[z]}$};
        \draw[axis] (-0.5,4,0) -- (4.5,4,0) node[right]{$\mathrm{Re}{[z]}$};
        \draw[upper] (0,4,0) -- (4,4,0) -- (4,4,3) -- (0,4,3) -- cycle;

        \draw[sub] (0.5,4,2) node[black, above]{$z_1$} -- (1.5,4,2) node[black, above]{$z_2$};
        \draw[sub] (2.5,4,2) node[black, above]{$z_3$} -- (3.5,4,2) node[black, above]{$z_4$};
        \draw[geo, variable=\t, domain=0:-1*pi, samples=200] plot({2+cos(\t r)*3/2},{4+sin(\t r)*3/2},{2});
        \draw[geo, variable=\t, domain=0:-1*pi, samples=200] plot({2+cos(\t r)/2},{4+sin(\t r)/2},{2});
    \end{tikzpicture}
    \end{tabular}}
    \caption{Disjoint subsystem $X \cup Y$ where the subsystems have a cross-ratio $|x|<1/2$ (left) and $|x|>1/2$ (right).}
    \label{fig:vacuum MI geodesics}
\end{figure}
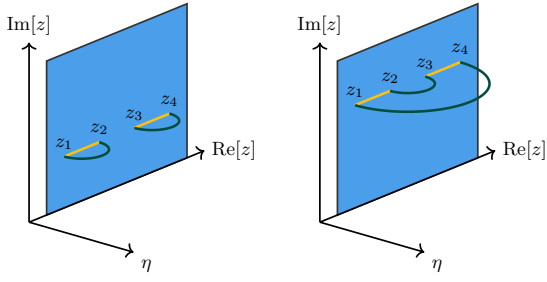

The s-channel expansion converges for $|x|<1$  and the t-channel for $|1-x| < 1$ giving us the result that there is a sharp transition at $x = 1/2$
\be 
S_{X\cup Y}=\min\left(\frac{c}{3}\ln\frac{x}{\epsilon},\frac{c}{3}\ln\frac{1-x}{\epsilon}\right) 
\ee
This can be nicely visualized with holography~\cite{headrick2010a}, as shown in Fig.~\ref{fig:vacuum MI geodesics}.
When the two subsystems are close together (have large cross-ratio), the minimal geodesic is the one that connects the pairs $[z_1,z_4]$ and $[z_2,z_3]$, when they are farther apart (have a small cross-ratio) it is the one that connects the pairs $[z_1,z_2]$ and $[z_3,z_4]$.
This perspective can be productively extended to the case with a boundary.
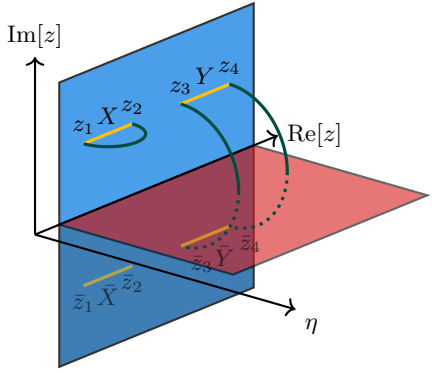
\begin{figure}
    \centering
    \tdplotsetmaincoords{70}{-50}
    \begin{tikzpicture}[
    tdplot_main_coords,
    axis/.style={->,black,thick},
    line/.style={very thick, color=black},
    upper/.style={thick, fill=color_upper, opacity=0.8},
    lower/.style={thick, fill=color_lower, opacity=0.8},
    boundary/.style={thick, fill=color_boundary, opacity=0.6},
    sub/.style={very thick, color=color_sub, opacity=1},
    geo/.style={very thick, color=color_geo, opacity=1}]
        
        \draw[axis] (-0.5,4,0) -- (-0.5,4,2.5) node[above]{$\mathrm{Im}{[z]}$};
        \draw[axis] (-0.5,4,0) -- (4.5,4,0) node[right]{$\mathrm{Re}{[z]}$};
        \draw[upper] (0,4,0) -- (4,4,0) -- (4,4,2) -- (0,4,2) -- cycle;
        \draw[lower] (0,4,0) -- (4,4,0) -- (4,4,-2) -- (0,4,-2)  -- cycle;
        \draw[boundary] (0,1,0) -- (4,1,0) -- (4,4,0) -- (0,4,0) -- cycle;
        \draw[sub] (0.5,4,1) node[black, above]{$z_1$} -- (1,4,1) node[black, above]{$X$} -- (1.5,4,1) node[black, above]{$z_2$};
        \draw[sub] (2.5,4,1) node[black, above]{$z_3$} -- (3,4,1) node[black, above]{$Y$} -- (3.5,4,1) node[black, above]{$z_4$};
        \draw[sub,opacity=0.6] (0.5,4,-1) node[black, below]{$\bar z_1$} -- (1,4,-1) node[black, below]{$\bar{X}$} -- (1.5,4,-1) node[black, below]{$\bar z_2$};
        \draw[sub,opacity=0.6] (2.5,4,-1) node[black, below right]{$\bar z_3$} -- (3,4,-1) node[black, below right]{$\bar{Y}$} -- (3.5,4,-1) node[black, below right]{$\bar z_4$};
        \draw[geo, variable=\t, domain=0:-1*pi, samples=200] plot({1+cos(\t r)/2},{4+sin(\t r)/2},{1});
        \draw[geo, variable=\t, domain=0:-1*pi/2, samples=200] plot({3.5},{4+sin(\t r)},{cos(\t r)});
        \draw[geo, variable=\t, domain=-1*pi/2:-2*pi/2, samples=200, dotted] plot({3.5},{4+sin(\t r)},{cos(\t r)});
        \draw[geo, variable=\t, domain=0:-1*pi/2, samples=200] plot({2.5},{4+sin(\t r)},{cos(\t r)});
        \draw[geo, variable=\t, domain=-1*pi/2:-2*pi/2, samples=200, dotted] plot({2.5},{4+sin(\t r)},{cos(\t r)});
        \draw[axis] (-0.5,4,0) -- (-0.5,-0.5,0) node[below right]{$\eta$};
    \end{tikzpicture}
    \caption{Upper half-plane with two subsystems $X=[z_1,z_2]$ and $Y=[z_3,z_4]$ and their mirror images $\bar{X}=[\bar{z}_2,\bar{z}_2]$ and $\bar{Y}=[\bar{z}_3,\bar{z}_4]$ on the lower half-plane. (Red: Conformal boundary separating upper and lower half-plane).}
    \label{fig:upper half-plane}
\end{figure}
Consider two subsystems $X = [z_1, z_2]$ and $Y = [z_3, z_4]$ on the upper half-plane, as shown in Fig.~\ref{fig:upper half-plane}.
Rather than deal with a complicated conformal block expansion (see Sec.~\ref{sec:section2}) we will use the AdS/BCFT formalism~\cite{fujita2011,takayanagi2011d,sully2021a} to visualize entanglement entropies.
The entanglement entropy of $X\cup Y$ is given by the minimal homologous geodesic~\cite{ryu2006b,ryu2006c}.
Different choices of boundary conditions on the BCFT will put different amounts of energy on the end-of-the-world brane causing it to jut into the bulk at different angles with respect to the upper half-plane, as captured in eq.~(\ref{eq:S-bdy}).

The figure shows our physical system ${\rm Im}(z)\geq 0,\eta = 0$ (light blue) as well as its holographic dual geometry $\eta>0$ (red) and a fictitious image of our system ${\rm Im}(z) < 0$ (dark blue).
The method of images is introduced here for the same reason we introduce image charges in electromagnetism: 
It is a convenient way of enforcing the appropriate boundary conditions.
As mentioned in Sec.~\ref{sec:section2} we can have geodesics that are attached to the subsystem endpoints (as you see for subsystem $X$) and those that are attached to the end-of-the-world brane (as you see for subsystem $Y$). The disconnected geodesics end on the brane; each is a fraction of the geodesic curve connecting the point $z_i$ to its mirror image $\bar{z}_i$, which is naturally parameterized by $k = (1+T)/(1-T)$.

The associated entanglement entropies are then
\begin{equation}
    \begin{aligned}
        S_{X}^\mathrm{con} &= \frac{c}{6} \ln\frac{|z_{12}||z_{\bar{1}\bar{2}}|}{\epsilon^2_\mathrm{UHP}} \\
        S_{Y}^\mathrm{dis} &= \frac{c}{6} \ln\frac{|z_{3 \bar 3}||z_{4 \bar 4}| k}{\epsilon^2_\mathrm{UHP}},
    \end{aligned}
\end{equation}
where $\epsilon_\mathrm{UHP}$ is a short-distance cutoff on the upper half-plane and we use the notation $z_{ij}=z_i-z_j$ and $z_{i\bar j}=z_i-\bar{z}_j$.
Under a conformal transformation it is related to $\epsilon_{\rm UV}$ by a Jacobian as
\begin{equation}
    \epsilon_\mathrm{UHP} = |f'(w)| \epsilon_\mathrm{UV},
\end{equation}
where $f'(w)$ is the derivative of the map to the UHP with respect to the complex coordinate $w = x + i\tau$.

If $X$ is very close to the boundary then the disconnected geodesic will be smaller than the connected one. 
Giving the entanglement entropy
\be S_{X}^\mathrm{dis} = \frac{c}{6} \ln\frac{|z_{1 \bar 1}||z_{2 \bar 2}| k}{\epsilon^2_\mathrm{UHP}}.\ee
The transition between $S_{x}^\mathrm{con}$ and $S_{x}^\mathrm{dis}$ is well established and for example nicely shown with a holographic computation in~\cite{shimaji2019a}.
One can find which $z_1,z_2$ the transition occurs at by setting $S_{X}^\mathrm{con}=S_{X}^\mathrm{dis}$ and solving.
We get
\be \frac{|z_{12}||z_{\bar{1}\bar{2}}|}{|z_{1 \bar 1}||z_{2 \bar 2}|}=k.\ee
Note that this takes the form of a cross-ratio\footnote{If one is used to $PSL(2,\mathbb{C})$ cross-ratios this can be written as $\sqrt{x\bar{x}}$ where $x=(z_{12}z_{\bar{1}\bar{2}})/(z_{1 \bar 1}z_{2 \bar 2})\in\mathbb{C}$}. As we will deal with many similar transitions we define the following class of cross-ratios
\be v_{ij} \equiv \frac{|z_{ij}||z_{\bar{i}\bar{j}}|}{|z_{i \bar i}||z_{j \bar j}|}.\label{eq:crossratio}\ee
In this notation, we say that if $v_{12}<k$ the subsystem $X$ will have entanglement entropy $S_{X}^\mathrm{con}$ and if $v_{12}>k$ it will have entanglement entropy $S_{X}^\mathrm{dis}$.
This is a useful way of thinking of things because the effect of the boundary is packaged nicely into the phase behavior, the cross-ratio is invariant under the global conformal group, and it does not pick up a Jacobian under conformal mapping, so such conditions have a nice pull-back to the real-time dynamics.\\

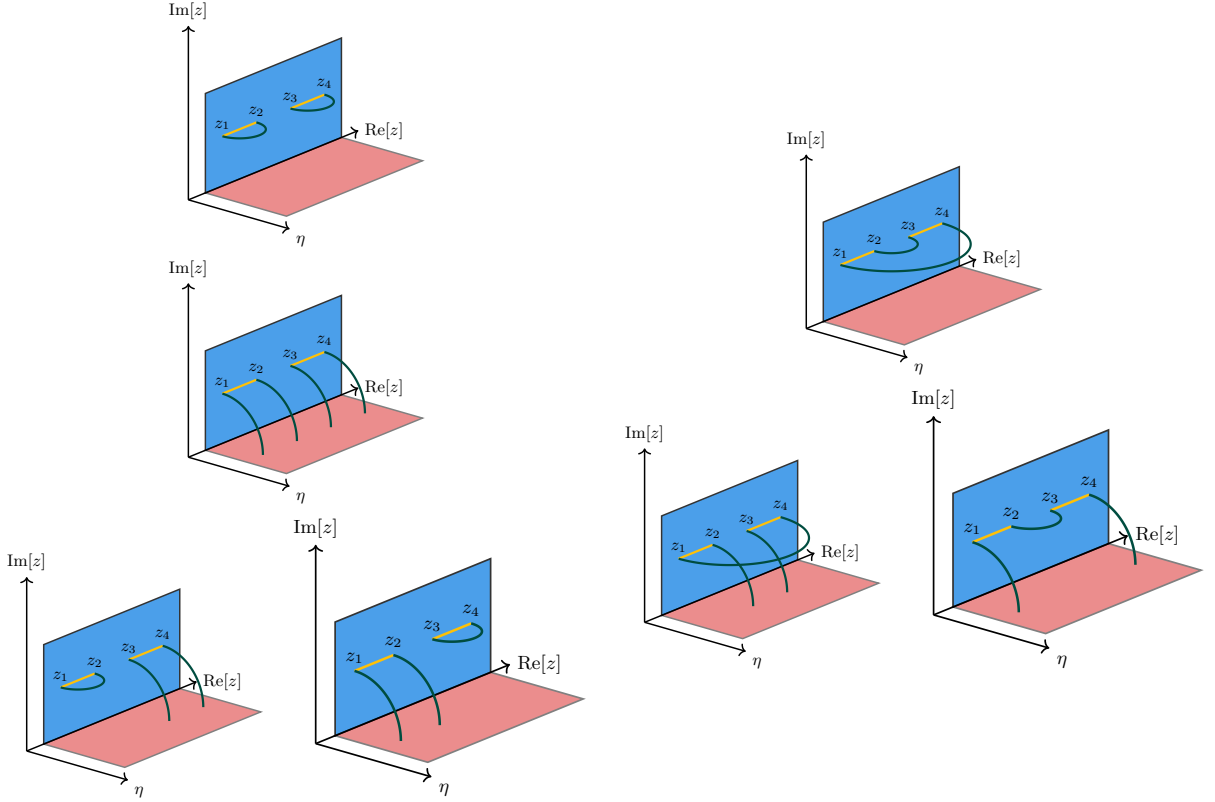
\begin{figure*}
    \centering
    \begin{minipage}{.45\linewidth}
    \centering
    \scalebox{.7}{\input{Plots/phase1.tex}}\\
    \scalebox{.7}{\input{Plots/phase4.tex}}\\
    \scalebox{.7}{\input{Plots/phase3a.tex}}\scalebox{.8}{\input{Plots/phase3b.tex}}
    \end{minipage} 
    \begin{minipage}{.45\linewidth}
    \centering
    
    \scalebox{.7}{\input{Plots/phase2.tex}}\\
    \scalebox{.7}{\input{Plots/phase5a.tex}}\scalebox{.8}{\input{Plots/phase5b.tex}}\\
    \end{minipage}

    \caption{On the left are the phases 1, 2, 3a and 3b respectively, and on the right are the phases 4, 5a and 5b.}
    \label{fig:Phases}
\end{figure*}

Having reviewed known phase structures, let us now turn to all expressions for the entanglement entropy of $X \cup Y$ that contribute\footnote{By naive combinatorics one expects 10 options (one can consider the number of ways to partition 4 labeled vertices into pairs and/or singletons). However 3 of the possibilities will never contribute as their geodesic length will always be strictly longer than at least one of the seven geodesics considered here.}
\begin{equation}
\label{eq:AuB phases}
    \begin{aligned}
        S^{(1)}_{X \cup Y} &= \frac{c}{6}\ln\frac{|z_{12}||z_{\bar 1 \bar 2}||z_{34}||z_{\bar 3 \bar 4}|}{\epsilon^4_\mathrm{UHP}} \\
        S^{(2)}_{X \cup Y} &= \frac{c}{6}\ln\frac{|z_{1 \bar 1}| |z_{2 \bar 2}| |z_{3\bar 3}| |z_{4 \bar 4}| k^2}{\epsilon^4_\mathrm{UHP}} \\
        S^{(3\mathrm{a})}_{X \cup Y} &= \frac{c}{6}\ln\frac{|z_{12}||z_{\bar 1 \bar 2}||z_{3 \bar 3}||z_{4 \bar 4}| k}{\epsilon^4_\mathrm{UHP}} \\
        S^{(3\mathrm{b})}_{X \cup Y} &= \frac{c}{6}\ln\frac{|z_{34}||z_{\bar 3 \bar 4}||z_{ 1 \bar 1}||z_{2 \bar 2}| k}{\epsilon^4_\mathrm{UHP}} \\
        \\
        \hline
        \\
        S^{(4)}_{X \cup Y} &= \frac{c}{6}\ln\frac{|z_{14}||z_{\bar 1 \bar 4}||z_{23}||z_{\bar 2 \bar 3}|}{\epsilon^4_\mathrm{UHP}} \\
        S^{(5\mathrm{a})}_{X \cup Y} &= \frac{c}{6}\ln\frac{|z_{14}| |z_{\bar 1 \bar 4}| |z_{2 \bar 2}| |z_{3 \bar 3}| k}{\epsilon^4_\mathrm{UHP}} \\
        S^{(5\mathrm{b})}_{X \cup Y} &= \frac{c}{6}\ln\frac{|z_{23}| |z_{\bar 2 \bar 3}| |z_{1 \bar 1}| |z_{4 \bar 4}| k}{\epsilon^4_\mathrm{UHP}}.
    \end{aligned}
\end{equation}
The above equations can be visualized holographically, as shown in Fig.~\ref{fig:Phases}.

Some clarification of labeling is in order. Since $S_{X\cup Y}$ is invariant under relabeling $X \leftrightarrow Y$ we label with $a/b$ the phases that are sent to each other under this permutation.
In addition, we separate into phases that are decomposable $S_{X\cup Y}=S_X+S_Y$ as they will not contribute to mutual information, these are the phases 1, 2, 3a and 3b\footnote{That means the phases of the individual subsystems $A$ and $B$ can be expressed as a phase of the disjoint subsystem $A \cup B$. The individual subsystems can each be either in the connected (C) or disconnected (D) phase, so together they have the four options (CC), (CD), (DC) and (DD), which are just the phases 1, 2, 3a and 3b of $A \cup B$. From here on we will label the phases of $A$ and $B$ via the phases of $A \cup B$. If necessary, we will express the total phase of the system as a tuple (Phase$_{A \cup B}$, Phase$_{A \& B}$).}.
These phases are invariant under a $D_4$ subgroup of the permutation of endpoints and the remaining phases 4, 5a and 5b are invariant under a different $D_4$ subgroup.
This is worked out in detail in Appendix C.
The transition between any two of the phases can be written in terms of cross-ratios as above, a tabulation of the various inequalities can be found in Appendix B. Combining the possible phases of $X\cup Y$ with those of the individual intervals, we find six distinct mutual-information phases. More explicitly, up to relabeling, there are two possible nontrivial phases for $X\cup Y$, and each can be paired with three inequivalent decomposable phases for $S_X$ and $S_Y$, as represented by phases 1, 2, 3a, and 3b in Fig.~\ref{fig:Phases}.

%% file: Plots/phase1.tex
\tdplotsetmaincoords{70}{-50}
        \begin{tikzpicture}[
        		tdplot_main_coords,
                axis/.style={->,black,thick},
                line/.style={very thick, color=black},
                upper/.style={thick, fill=color_upper, opacity=0.8},
                lower/.style={thick, fill=color_lower, opacity=0.8},
                boundary/.style={thick, fill=color_boundary, opacity=0.5},
                sub/.style={very thick, color=color_sub, opacity=1},
                geo/.style={very thick, color=color_geo, opacity=1}]            
            \draw[axis] (-0.5,4,0) -- (-0.5,1.5,0) node[below right]{$\eta$};
            \draw[axis] (-0.5,4,0) -- (-0.5,4,3.5) node[above]{$\mathrm{Im}{[z]}$};
            \draw[axis] (-0.5,4,0) -- (4.5,4,0) node[right]{$\mathrm{Re}{[z]}$};
            \draw[upper] (0,4,0) -- (4,4,0) -- (4,4,2) -- (0,4,2) -- cycle;
            \draw[boundary] (0,2,0) -- (4,2,0) -- (4,4,0) -- (0,4,0) -- cycle;
            \draw[sub] (0.5,4,1) node[black, above]{$z_1$} -- (1.5,4,1) node[black, above]{$z_2$};
            \draw[sub] (2.5,4,1) node[black, above]{$z_3$} -- (3.5,4,1) node[black, above]{$z_4$};
            \draw[geo, variable=\t, domain=0:-1*pi, samples=200] plot({1+cos(\t r)/2},{4+sin(\t r)/2},{1});
            \draw[geo, variable=\t, domain=0:-1*pi, samples=200] plot({3+cos(\t r)/2},{4+sin(\t r)/2},{1});
        \end{tikzpicture}
    

%% file: Plots/phase4.tex
\tdplotsetmaincoords{70}{-50}
\begin{tikzpicture}[
        		tdplot_main_coords,
                axis/.style={->,black,thick},
                line/.style={very thick, color=black},
                upper/.style={thick, fill=color_upper, opacity=0.8},
                lower/.style={thick, fill=color_lower, opacity=0.8},
                boundary/.style={thick, fill=color_boundary, opacity=0.5},
                sub/.style={very thick, color=color_sub, opacity=1},
                geo/.style={very thick, color=color_geo, opacity=1}]
            
            \draw[axis] (-0.5,4,0) -- (-0.5,1.5,0) node[below right]{$\eta$};
            \draw[axis] (-0.5,4,0) -- (-0.5,4,3.5) node[above]{$\mathrm{Im}{[z]}$};
            \draw[axis] (-0.5,4,0) -- (4.5,4,0) node[right]{$\mathrm{Re}{[z]}$};
            \draw[upper] (0,4,0) -- (4,4,0) -- (4,4,2) -- (0,4,2) -- cycle;
            \draw[boundary] (0,2,0) -- (4,2,0) -- (4,4,0) -- (0,4,0) -- cycle;
            \draw[sub] (0.5,4,1) node[black, above]{$z_1$} -- (1.5,4,1) node[black, above]{$z_2$};
            \draw[sub] (2.5,4,1) node[black, above]{$z_3$} -- (3.5,4,1) node[black, above]{$z_4$};
            \draw[geo, variable=\t, domain=0:-1*pi/2, samples=200] plot({.5},{4+sin(\t r)},{cos(\t r)});
            \draw[geo, variable=\t, domain=0:-1*pi/2, samples=200] plot({1.5},{4+sin(\t r)},{cos(\t r)});
            \draw[geo, variable=\t, domain=0:-1*pi/2, samples=200] plot({3.5},{4+sin(\t r)},{cos(\t r)});
            \draw[geo, variable=\t, domain=0:-1*pi/2, samples=200] plot({2.5},{4+sin(\t r)},{cos(\t r)});
        \end{tikzpicture}

%% file: Plots/phase3a.tex
        \tdplotsetmaincoords{70}{-50}
        \begin{tikzpicture}[
        		tdplot_main_coords,
                axis/.style={->,black,thick},
                line/.style={very thick, color=black},
                upper/.style={thick, fill=color_upper, opacity=0.8},
                lower/.style={thick, fill=color_lower, opacity=0.8},
                boundary/.style={thick, fill=color_boundary, opacity=0.5},
                sub/.style={very thick, color=color_sub, opacity=1},
                geo/.style={very thick, color=color_geo, opacity=1}]
            
            \draw[axis] (-0.5,4,0) -- (-0.5,1.5,0) node[below right]{$\eta$};
            \draw[axis] (-0.5,4,0) -- (-0.5,4,3.5) node[above]{$\mathrm{Im}{[z]}$};
            \draw[axis] (-0.5,4,0) -- (4.5,4,0) node[right]{$\mathrm{Re}{[z]}$};
            \draw[upper] (0,4,0) -- (4,4,0) -- (4,4,2) -- (0,4,2) -- cycle;
            \draw[boundary] (0,2,0) -- (4,2,0) -- (4,4,0) -- (0,4,0) -- cycle;
            \draw[sub] (0.5,4,1) node[black, above]{$z_1$} -- (1.5,4,1) node[black, above]{$z_2$};
            \draw[sub] (2.5,4,1) node[black, above]{$z_3$} -- (3.5,4,1) node[black, above]{$z_4$};
            \draw[geo, variable=\t, domain=0:-1*pi, samples=200] plot({1+cos(\t r)/2},{4+sin(\t r)/2},{1});
            \draw[geo, variable=\t, domain=0:-1*pi/2, samples=200] plot({3.5},{4+sin(\t r)},{cos(\t r)});
            \draw[geo, variable=\t, domain=0:-1*pi/2, samples=200] plot({2.5},{4+sin(\t r)},{cos(\t r)});
        \end{tikzpicture}

    

%% file: Plots/phase3b.tex
   \tdplotsetmaincoords{70}{-50}
        \begin{tikzpicture}[
        		tdplot_main_coords,
                axis/.style={->,black,thick},
                line/.style={very thick, color=black},
                upper/.style={thick, fill=color_upper, opacity=0.8},
                lower/.style={thick, fill=color_lower, opacity=0.8},
                boundary/.style={thick, fill=color_boundary, opacity=0.5},
                sub/.style={very thick, color=color_sub, opacity=1},
                geo/.style={very thick, color=color_geo, opacity=1}]
            
            \draw[axis] (-0.5,4,0) -- (-0.5,1.5,0) node[below right]{$\eta$};
            \draw[axis] (-0.5,4,0) -- (-0.5,4,3.5) node[above]{$\mathrm{Im}{[z]}$};
            \draw[axis] (-0.5,4,0) -- (4.5,4,0) node[right]{$\mathrm{Re}{[z]}$};
            \draw[upper] (0,4,0) -- (4,4,0) -- (4,4,2) -- (0,4,2) -- cycle;
            \draw[boundary] (0,2,0) -- (4,2,0) -- (4,4,0) -- (0,4,0) -- cycle;
            \draw[sub] (0.5,4,1) node[black, above]{$z_1$} -- (1.5,4,1) node[black, above]{$z_2$};
            \draw[sub] (2.5,4,1) node[black, above]{$z_3$} -- (3.5,4,1) node[black, above]{$z_4$};
            \draw[geo, variable=\t, domain=0:-1*pi, samples=200] plot({3+cos(\t r)/2},{4+sin(\t r)/2},{1});
            \draw[geo, variable=\t, domain=0:-1*pi/2, samples=200] plot({.5},{4+sin(\t r)},{cos(\t r)});
            \draw[geo, variable=\t, domain=0:-1*pi/2, samples=200] plot({1.5},{4+sin(\t r)},{cos(\t r)});
        \end{tikzpicture}

%% file: Plots/phase2.tex
\tdplotsetmaincoords{70}{-50}
\begin{tikzpicture}[
        		tdplot_main_coords,
                axis/.style={->,black,thick},
                line/.style={very thick, color=black},
                upper/.style={thick, fill=color_upper, opacity=0.8},
                lower/.style={thick, fill=color_lower, opacity=0.8},
                boundary/.style={thick, fill=color_boundary, opacity=0.5},
                sub/.style={very thick, color=color_sub, opacity=1},
                geo/.style={very thick, color=color_geo, opacity=1}]
            
            \draw[axis] (-0.5,4,0) -- (-0.5,1.5,0) node[below right]{$\eta$};
            \draw[axis] (-0.5,4,0) -- (-0.5,4,3.5) node[above]{$\mathrm{Im}{[z]}$};
            \draw[axis] (-0.5,4,0) -- (4.5,4,0) node[right]{$\mathrm{Re}{[z]}$};
            \draw[upper] (0,4,0) -- (4,4,0) -- (4,4,2) -- (0,4,2) -- cycle;
            \draw[boundary] (0,2,0) -- (4,2,0) -- (4,4,0) -- (0,4,0) -- cycle;
            \draw[sub] (0.5,4,1) node[black, above]{$z_1$} -- (1.5,4,1) node[black, above]{$z_2$};
            \draw[sub] (2.5,4,1) node[black, above]{$z_3$} -- (3.5,4,1) node[black, above]{$z_4$};
            \draw[geo, variable=\t, domain=0:-1*pi, samples=200] plot({2+cos(\t r)*3/2},{4+sin(\t r)*3/2},{1});
            \draw[geo, variable=\t, domain=0:-1*pi, samples=200] plot({2+cos(\t r)/2},{4+sin(\t r)/2},{1});
        \end{tikzpicture}

%% file: Plots/phase5a.tex
\tdplotsetmaincoords{70}{-50}
        \begin{tikzpicture}[
        		tdplot_main_coords,
                axis/.style={->,black,thick},
                line/.style={very thick, color=black},
                upper/.style={thick, fill=color_upper, opacity=0.8},
                lower/.style={thick, fill=color_lower, opacity=0.8},
                boundary/.style={thick, fill=color_boundary, opacity=0.5},
                sub/.style={very thick, color=color_sub, opacity=1},
                geo/.style={very thick, color=color_geo, opacity=1}]
            
            \draw[axis] (-0.5,4,0) -- (-0.5,1.5,0) node[below right]{$\eta$};
            \draw[axis] (-0.5,4,0) -- (-0.5,4,3.5) node[above]{$\mathrm{Im}{[z]}$};
            \draw[axis] (-0.5,4,0) -- (4.5,4,0) node[right]{$\mathrm{Re}{[z]}$};
            \draw[upper] (0,4,0) -- (4,4,0) -- (4,4,2) -- (0,4,2) -- cycle;
            \draw[boundary] (0,2,0) -- (4,2,0) -- (4,4,0) -- (0,4,0) -- cycle;
            \draw[sub] (0.5,4,1) node[black, above]{$z_1$} -- (1.5,4,1) node[black, above]{$z_2$};
            \draw[sub] (2.5,4,1) node[black, above]{$z_3$} -- (3.5,4,1) node[black, above]{$z_4$};
            \draw[geo, variable=\t, domain=0:-1*pi, samples=200] plot({2+cos(\t r)*3/2},{4+sin(\t r)*3/2},{1});
            \draw[geo, variable=\t, domain=0:-1*pi/2, samples=200] plot({1.5},{4+sin(\t r)},{cos(\t r)});
            \draw[geo, variable=\t, domain=0:-1*pi/2, samples=200] plot({2.5},{4+sin(\t r)},{cos(\t r)});
        \end{tikzpicture}

     

%% file: Plots/phase5b.tex
 \tdplotsetmaincoords{70}{-50}
        \begin{tikzpicture}[
        		tdplot_main_coords,
                axis/.style={->,black,thick},
                line/.style={very thick, color=black},
                upper/.style={thick, fill=color_upper, opacity=0.8},
                lower/.style={thick, fill=color_lower, opacity=0.8},
                boundary/.style={thick, fill=color_boundary, opacity=0.5},
                sub/.style={very thick, color=color_sub, opacity=1},
                geo/.style={very thick, color=color_geo, opacity=1}]
            
            \draw[axis] (-0.5,4,0) -- (-0.5,1.5,0) node[below right]{$\eta$};
            \draw[axis] (-0.5,4,0) -- (-0.5,4,3.5) node[above]{$\mathrm{Im}{[z]}$};
            \draw[axis] (-0.5,4,0) -- (4.5,4,0) node[right]{$\mathrm{Re}{[z]}$};
            \draw[upper] (0,4,0) -- (4,4,0) -- (4,4,2) -- (0,4,2) -- cycle;
            \draw[boundary] (0,2,0) -- (4,2,0) -- (4,4,0) -- (0,4,0) -- cycle;
            \draw[sub] (0.5,4,1) node[black, above]{$z_1$} -- (1.5,4,1) node[black, above]{$z_2$};
            \draw[sub] (2.5,4,1) node[black, above]{$z_3$} -- (3.5,4,1) node[black, above]{$z_4$};
            \draw[geo, variable=\t, domain=0:-1*pi, samples=200] plot({2+cos(\t r)/2},{4+sin(\t r)/2},{1});
            \draw[geo, variable=\t, domain=0:-1*pi/2, samples=200] plot({.5},{4+sin(\t r)},{cos(\t r)});
            \draw[geo, variable=\t, domain=0:-1*pi/2, samples=200] plot({3.5},{4+sin(\t r)},{cos(\t r)});
        \end{tikzpicture}

%% file: Sections/4_Dynamical_Phase_Transitions.tex
As explained in the previous section, the entanglement entropy of two disjoint intervals on the UHP has a variety of phases. Which phase a given configuration is in is determined by a finite set of competing geodesic configurations (or equivalently a dominant conformal block). The competition between two phases can be quantified in terms the value of conformally invariant cross-ratios of the endpoints. Among the seven phases listed in eq.~\eqref{eq:AuB phases}, only phases 4, 5a and 5b are non-decomposable,
\be
S_{X\cup Y}\neq S_X+S_Y,
\ee
and therefore only these phases give non-vanishing mutual information.

We now use this static UHP phase structure to describe real-time dynamics in CFTs with one conformal boundary. In particular, we focus on local splitting and joining quenches of a $(1+1)$-dimensional CFT on the infinite line.\footnote{From here on, intervals and coordinates on the world-sheet will be denoted by $A=[x_1,x_2]$, $B=[x_3,x_4]$, and $w=x+i\tau$, which after Wick rotation becomes $\tilde w=x-t$. Intervals and coordinates on the UHP will be denoted by $X=[z_1,z_2]$ and $Y=[z_3,z_4]$, as in Sec.~\ref{sec:section3}.}

The prescription is the same for all quench geometries with one conformal boundary. First, we map the Euclidean world-sheet with coordinate $w$ to the UHP by a conformal transformation
\be
z=f(w).
\ee
Second, we evaluate the UHP geodesic lengths for the image intervals
\be
X=[z_1,z_2]=[f(w_1),f(w_2)],\qquad
Y=[z_3,z_4]=[f(w_3),f(w_4)].
\ee
After Wick rotation, the endpoints $w_i$ become time dependent, and therefore so do the UHP coordinates $z_i$ and the cross-ratios $v_{ij}$.

For example, consider phase 1 of the disjoint interval $X\cup Y$. On the UHP its entropy is
\be
    S^{(1)}_{X \cup Y}
    =
    \frac{c}{6}
    \ln
    \frac{
    |z_{12}||z_{\bar 1 \bar 2}|
    |z_{34}||z_{\bar 3 \bar 4}|
    }{\epsilon^4_\mathrm{UHP}} .
\ee
Pulling this expression back to the physical world-sheet gives
\be
    \label{eq:mapped EE}
    S^{(1)}_{A \cup B}
    =
    \frac{c}{6}
    \ln
    \frac{
    |f_{12}||f_{\bar 1 \bar 2}|
    |f_{34}||f_{\bar 3 \bar 4}|
    }{
    |f'(w_1)||f'(w_2)||f'(w_3)||f'(w_4)|
    \epsilon^4_\mathrm{UV}
    },
\ee
where $f_{i\bar j}=f(w_i)-\bar f(w_j)$. We used that the cutoff  $\epsilon_{UHP}$ transforms with the Jacobian of the conformal map, as described in Sec.~\ref{sec:section3} in order to express our quantities in the physical cutoff $\epsilon_{UV}$. The same pull-back can be applied to all seven UHP phases. Thus the dynamical problem reduces to tracking which UHP phase minimizes the entropy as the time-dependent cross-ratios evolve.

\subsection{Splitting Quench}

We first consider a local splitting quench of a CFT on the infinite line. A convenient map from the world-sheet to the UHP is
\be
    f(w)
    =
    i \sqrt{\frac{w+i a}{w-i a}},
\label{eq:UHPmap-Vac}
\ee
where $w=x+i\tau$ and $a$ is the quench regulator as described in Section ~\ref{sec:section2}. The corresponding world-sheet is shown in the first column of Fig.~\ref{fig:Worldsheets}; this map, together with the other maps used below, is collected in Appendix~\ref{app:maps}.
For simplicity's sake, in what follows we set the boundary entropy to zero, $S_\mathrm{bdy}=0$, equivalently $T=0$ and $k=1$.

\begin{figure}[H]
    \centering
    \includegraphics[width=1\linewidth]{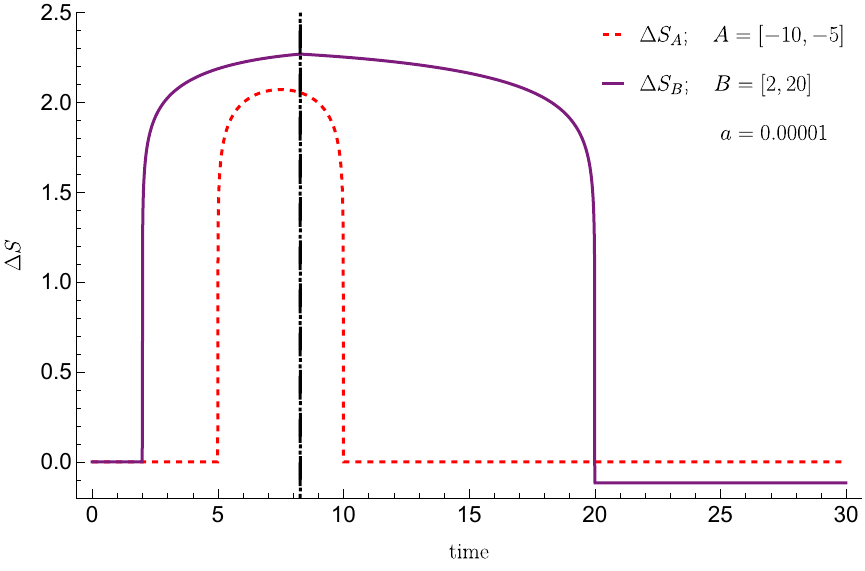}
    \caption{Entanglement entropy differences $\Delta S=S(t)-S(0)$ after a single splitting quench for contiguous subsystems $A$ and $B$. The dot-dashed black line marks the non-analytic point in $\Delta S_B$, where a phase transition occurs (there is no such point in $\Delta S_A$ for this setup).}
    \label{fig:splitting EE A}
\end{figure}

We focus on two intervals $A=[x_1,x_2]$ and $B=[x_3,x_4]$ placed on opposite sides of the splitting point. For a single interval on one side of the quench, the entanglement entropy develops a characteristic bump: it rises when the first signal from the quench reaches the interval and decreases when the second endpoint is reached. This behavior is visible in Fig.~\ref{fig:splitting EE A} and is often described using the quasiparticle picture, in which the quench emits entangled pairs of quasiparticles moving in opposite directions at the speed of light.

For sufficiently small intervals far from the quench, this bump is analytic between its initial rise and final decay; in this regime the connected geodesic remains minimal throughout the evolution. This is illustrated by the red curve in Fig.~\ref{fig:splitting EE A}. The orange curve shows that this is not generic. For the interval $B$, the entropy has a non-analytic peak, marked by the dashed gray line. At this time the minimal geodesic changes discontinuously from the connected to the disconnected configuration. In the UHP language this is precisely the transition from $v_{34}<k$ to $v_{34}>k$.

The quasiparticle picture correctly predicts the causal times at which the entropy begins to change, but it does not account for such non-analytic peaks. These features are instead naturally explained by the UHP phase structure. This becomes especially important for the mutual information.

\begin{figure}[H]
    \centering
    \includegraphics[width=\linewidth]{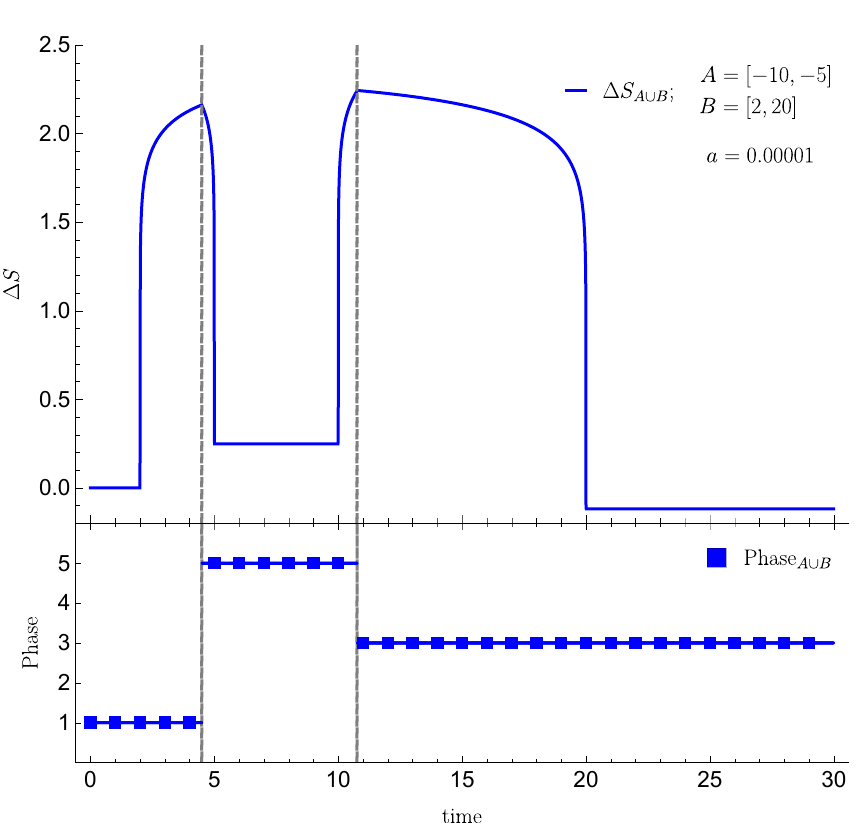}
    \caption{Top: Entanglement entropy difference $\Delta S=S(t)-S_0$ after a single splitting quench for the disjoint subsystem $A\cup B$, where $S_0$ is the entropy of the unsplit system at $t<0$. Dashed gray lines mark non-analytic points at which phase transitions occur in the phase of $A \cup B$, the dot-dashed black line marks the phase transition in $A \&B $. Bottom: Phase of $A\cup B$, using the labels of Sec.~\ref{sec:section3}.}
    \label{fig:splitting EE AuB}
\end{figure}

Let us now consider the disjoint interval $A\cup B$. As shown in Sec.~\ref{sec:section3}, the corresponding UHP interval $X\cup Y$ can be in any of the seven phases listed in eq.~\eqref{eq:AuB phases}. We therefore expect non-analyticities whenever the time-dependent cross-ratios cross one of the phase boundaries. An example is shown in Fig.~\ref{fig:splitting EE AuB}.

The main rises and drops in $S_{A\cup B}$ again occur at the causal times expected from the quasiparticle picture. However, the additional non-analytic peaks are phase transitions between different UHP geodesic configurations. In the example shown, the subsystem begins in phase 1 and subsequently transitions through phase 5 into phase 3.

These phase transitions have direct consequences for the mutual information,
\be
I_{A:B}=S_A+S_B-S_{A\cup B}.
\ee
We find two characteristic effects.

\begin{figure}[H]
    \centering
    \includegraphics[width=\linewidth]{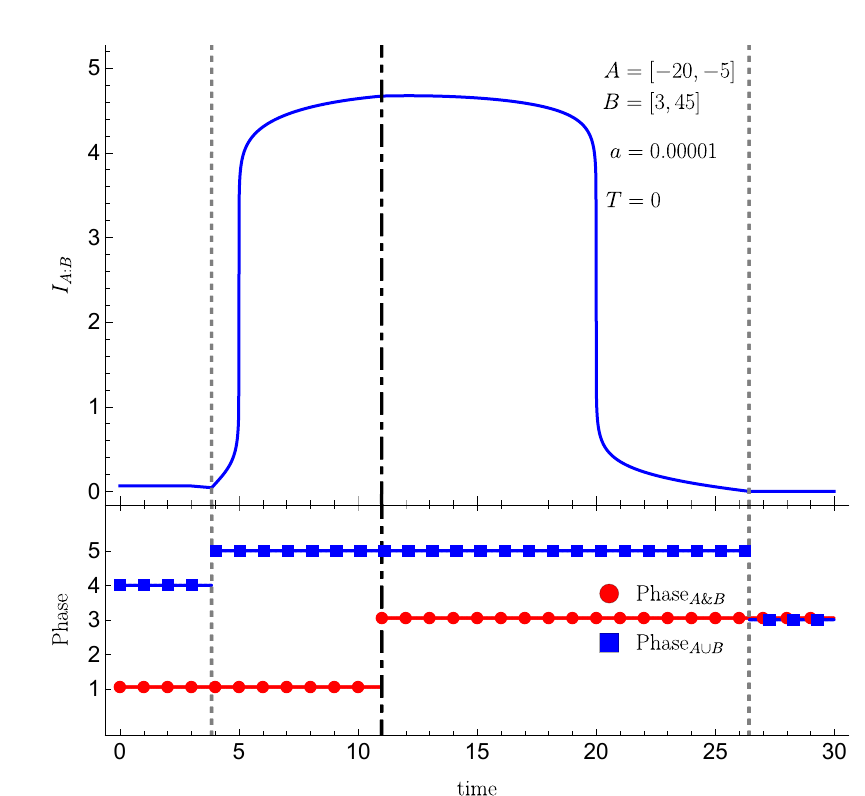}
    \caption{Panel 1: Mutual information after a single splitting quench. Gray dashed lines mark phase transitions of $A\cup B$, while black dot-dashed lines mark phase transitions of the individual subsystems $A$ or $B$. Panel 2: Phases of the disjoint subsystem $A\cup B$ and of the individual subsystems $A$ and $B$, where the latter are encoded by phases 1, 2, 3a and 3b of $A\cup B$.}
    \label{fig:splitting MI}
\end{figure}

First, the support of the mutual information is not determined solely by the quasiparticle light-cone times. In Fig.~\ref{fig:splitting MI}, a quasiparticle argument would predict that the mutual information returns to zero at $t=20$. Instead, the mutual information develops a tail and vanishes only at a later time, approximately $t\simeq 26.4$. A similar tail appears at the initial rise. The beginning and end of these tails coincide with phase transitions of the disjoint subsystem $A\cup B$.

The origin of these tails is simple in the UHP description. The phase of $X\cup Y$ is determined by the cross-ratio inequalities listed in Appendix~\ref{app:table of phase conditions}. In the static analysis of Sec.~\ref{sec:section3}, these inequalities depend only on the fixed endpoint coordinates $z_i$. After pulling back to the world-sheet, however, the same quantities become time dependent:
\be
v_{ij}\rightarrow v_{ij}(t).
\ee
Thus conditions such as
\be
\frac{v_{14}(t)}{v_{12}(t)v_{34}(t)}<k
\ee
become dynamical phase conditions. A phase transition occurs when one of these inequalities is saturated.

For example, suppose the system initially lies in phase 4, so that $I_{A:B}>0$. The possible phase boundaries out of phase 4 are
\begin{itemize}
    \item $\mathrm{Phase}~4 \rightarrow \mathrm{Phase}~1$:
    $\displaystyle \frac{v_{23}(t)v_{14}(t)}{v_{12}(t)v_{34}(t)}=1$,
    after which $I_{A:B}=0$.

    \item $\mathrm{Phase}~4 \rightarrow \mathrm{Phase}~3a$:
    $\displaystyle \frac{v_{23}(t)v_{14}(t)}{v_{12}(t)}=1$,
    after which $I_{A:B}=0$.

    \item $\mathrm{Phase}~4 \rightarrow \mathrm{Phase}~3b$:
    $\displaystyle \frac{v_{23}(t)v_{14}(t)}{v_{34}(t)}=1$,
    after which $I_{A:B}=0$.

    \item $\mathrm{Phase}~4 \rightarrow \mathrm{Phase}~2$:
    $\displaystyle v_{23}(t)v_{14}(t)=1$,
    after which $I_{A:B}=0$.

    \item $\mathrm{Phase}~4 \rightarrow \mathrm{Phase}~5a$:
    $\displaystyle v_{23}(t)=1$,
    after which $I_{A:B}>0$.

    \item $\mathrm{Phase}~4 \rightarrow \mathrm{Phase}~5b$:
    $\displaystyle v_{14}(t)=1$,
    after which $I_{A:B}>0$.
\end{itemize}
For $k=1$, the transition point is located at unity for all these conditions. The system remains in phase 4 as long as all relevant quantities remain on the phase-4 side of their inequalities. Once one of them crosses its transition value, the minimal geodesic changes and the system enters the corresponding new phase. The subsequent evolution is then governed by the set of inequalities appropriate to that phase.

\begin{figure}
    \centering
    \includegraphics[width=.8\linewidth]{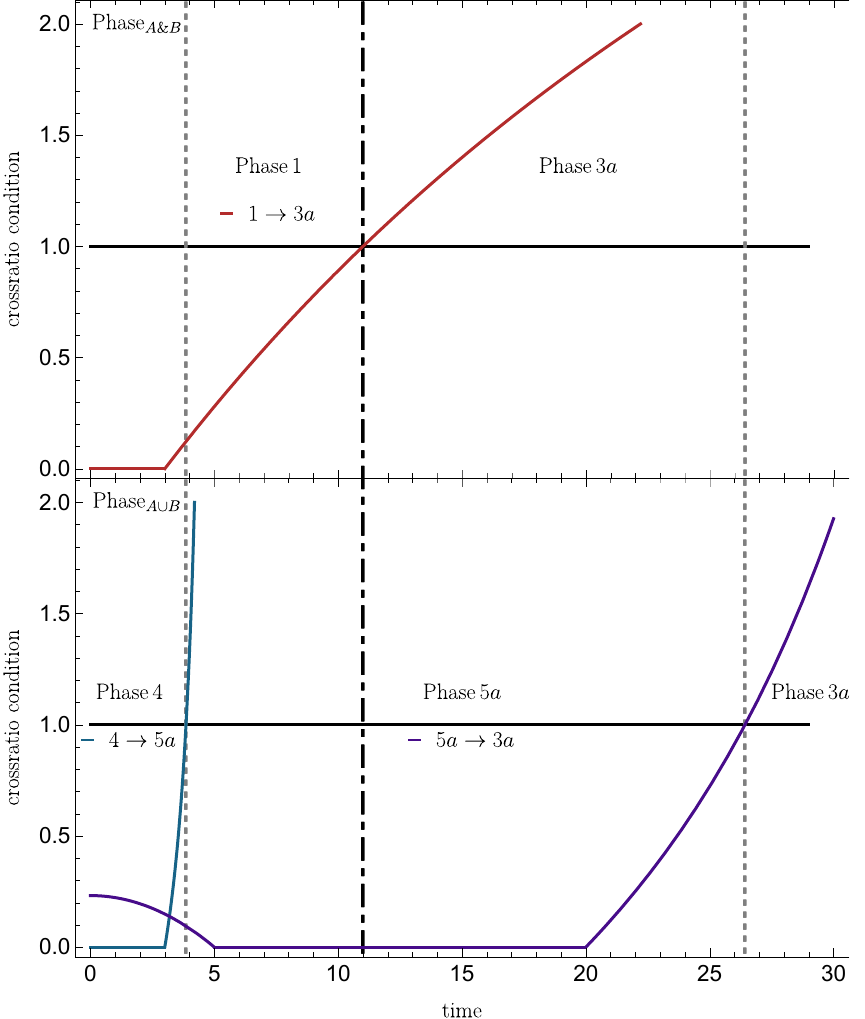}
    \caption{Panel 1: Time dependence of the phase-changing cross-ratio conditions for the individual intervals $A$ and $B$. Panel 2: Time dependence of the corresponding phase conditions for $A\cup B$. The horizontal black line marks the transition value for $k=1$.}
    \label{fig:splitting CR}
\end{figure}

This is shown in Fig.~\ref{fig:splitting CR}. The first panel displays the phase-changing cross-ratio conditions associated with the individual intervals $A$ and $B$. These are represented using phases 1, 2, 3a and 3b of the disjoint interval, since these phases encode whether the individual intervals are connected or disconnected, as explained in Sec.~\ref{sec:section3}. The second panel shows the relevant phase conditions for $A\cup B$. The crossings in this second panel are the transitions responsible for the mutual-information tails.

In the example of Fig.~\ref{fig:splitting MI}, the initial configuration has non-zero mutual information and hence begins in one of the non-decomposable phases. At the first transition, marked by the leftmost gray dashed line, the system enters phase 5 and the first tail begins. At the final gray dashed line the reverse mechanism terminates the tail, although the system ultimately settles into phase 3 rather than returning to phase 4.

\begin{figure}[H]
    \centering
    \includegraphics[width=\linewidth]{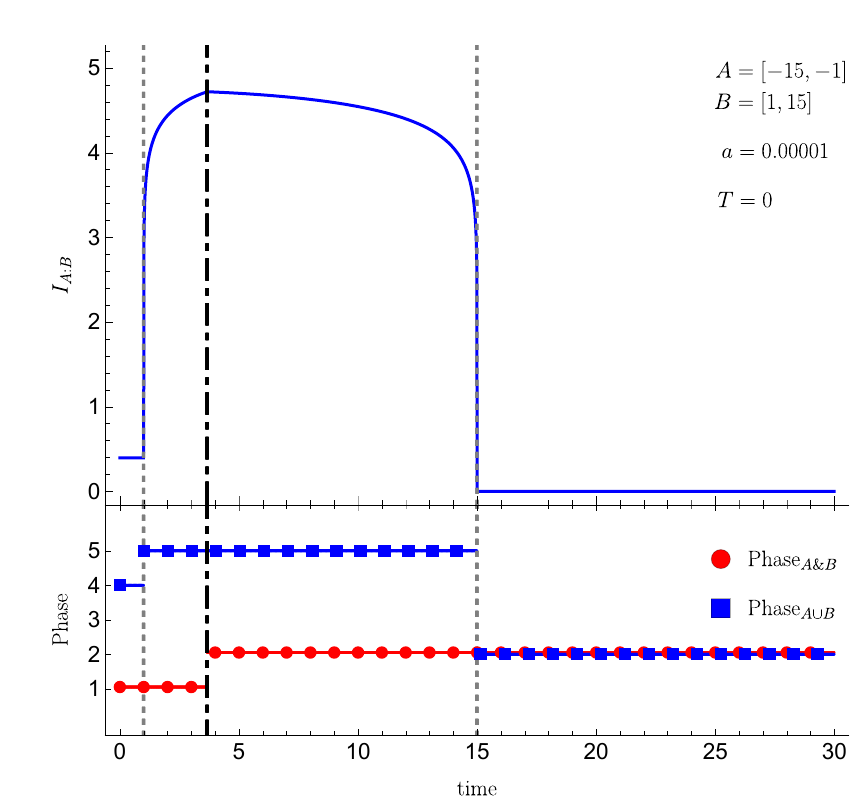}
    \caption{Top: Mutual information after a single splitting quench. Gray dashed lines mark phase transitions of $A\cup B$, while black dot-dashed lines mark phase transitions of the individual subsystems $A$ or $B$. Bottom: Phases of $A\cup B$ and of the individual subsystems, encoded as in Fig.~\ref{fig:splitting MI}.}
    \label{fig:splitting MI peak}
\end{figure}

Second, the mutual information can itself develop non-analytic peaks. These are caused not by a phase transition of $A\cup B$, but by a transition in one or both individual intervals. In Fig.~\ref{fig:splitting MI}, this is marked by the black dashed line. There one of the two intervals changes from the connected to the disconnected phase while the disjoint interval remains in a mutual-information phase. In Fig.~\ref{fig:splitting MI peak}, the setup is symmetric about the quench point, so both intervals undergo this transition simultaneously. The resulting non-analyticity in the mutual-information bump is correspondingly more pronounced.

In this symmetric example no visible tails appear, but the gray dashed lines still mark dynamical phase transitions of $A\cup B$. Thus even when the mutual information is supported on the quasiparticle time window, its detailed time dependence remains controlled by the UHP phase structure.

\subsection{Joining Quench}

The joining quench is treated in the same way. The Euclidean world-sheet is mapped to the UHP by
\be
    f(w)
    =
    i \sqrt{\frac{i a-w}{i a+w}},
\ee
where $w=x+i\tau$ and $a$ is the regulator. Before the quench, the system consists of two decoupled half-lines. They are joined at $t=0$, after which the system evolves as a CFT on the infinite line. Since the two halves are initially unentangled, the mutual information between intervals on opposite sides of the joining point vanishes at $t=0$.

\begin{figure}[H]
    \centering
    \includegraphics[width=\linewidth]{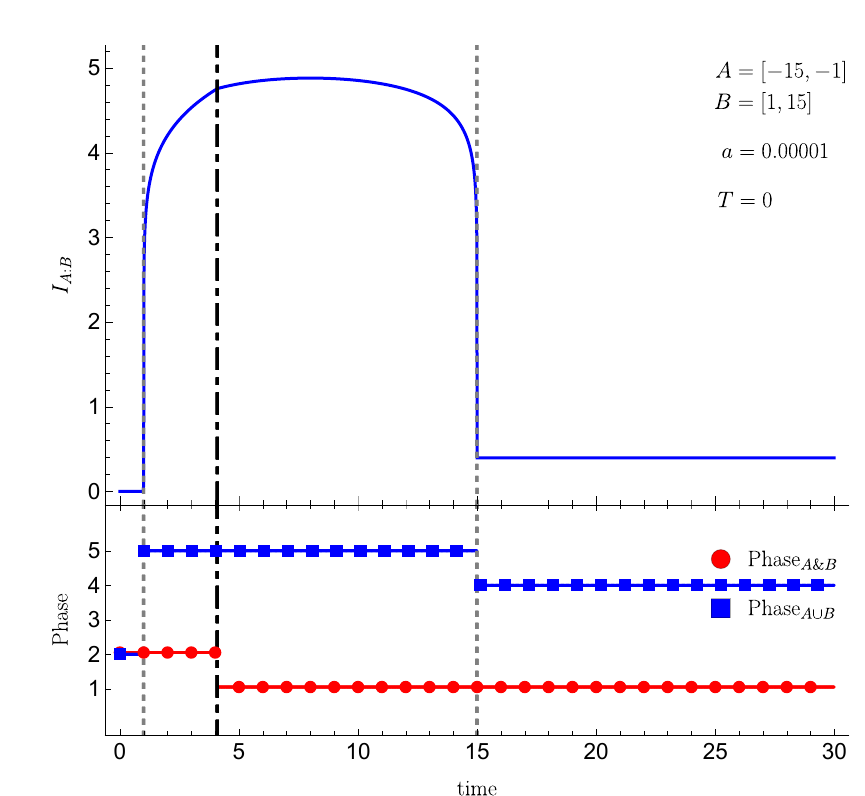}
    \caption{Top: Mutual information after a joining quench. Gray dashed lines mark phase transitions of $A\cup B$, while the black dot-dashed line marks a phase transition of the individual subsystems. Bottom: Phases of $A\cup B$ and of the individual subsystems, encoded by phases 1, 2, 3a and 3b as in Sec.~\ref{sec:section3}.}
    \label{fig:joining quench dynamics}
\end{figure}

The entropies of $A$, $B$ and $A\cup B$ are again obtained by pulling back the UHP expressions through the conformal map. The resulting dynamics is analogous to the splitting case. We find both types of dynamical phase transition: non-analytic peaks caused by transitions of the individual intervals, and tails caused by transitions of the disjoint interval $A\cup B$. These features are shown in Fig.~\ref{fig:joining quench dynamics}. As before, gray dashed lines indicate phase transitions of $A\cup B$, while the black dashed line indicates a transition of the individual intervals.

An interesting relation between the two protocols is that the joining quench evolves between phase configurations opposite to those of the splitting quench. The splitting system can begin with non-zero mutual information and relax to a configuration with zero mutual information. By contrast, the joining system starts with zero mutual information, because the two half-lines are initially decoupled, but it can evolve to a configuration with finite late-time mutual information.

\subsection{Vacuum versus Thermal Splitting}

As reviewed in Sec.~\ref{sec:section2} we can consider a state in a thermal ensemble by considering the system on a thermal circle. After a splitting or joining quench the geometry again contains one conformal boundary, and can therefore be mapped to the UHP. For a finite-temperature splitting quench, a convenient map is
\be
z(w)
=
i\sqrt{
\frac{e^{2\pi w/\beta}+i e^{2\pi a/\beta}}
     {e^{2\pi w/\beta}-i e^{2\pi a/\beta}}
},
\label{eq:UHPmap-T}
\ee
where $\beta$ is the inverse temperature and $a$ is the regulator. The same UHP phase prescription therefore applies.

\begin{figure}[H]
    \centering
    \includegraphics[width=\linewidth]{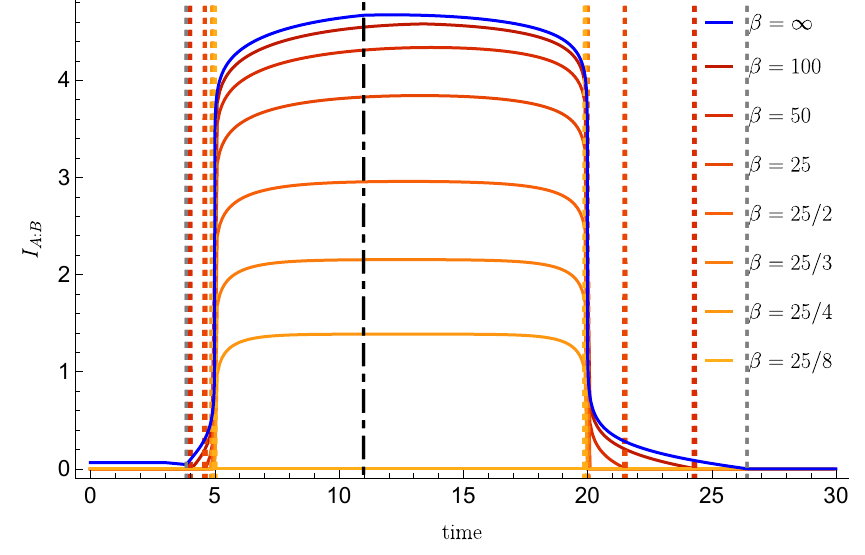}
    \caption{Mutual information after a finite-temperature splitting quench for several inverse temperatures. The blue (top most) curve shows the vacuum result, the curves below correspond to decreasing $\beta$. The gray and black dashed lines mark the phase transitions of the vacuum configuration shown in Fig.~\ref{fig:splitting MI}. The innermost colored dashed lines correspond to the lowest $\beta$ value and vice versa.}
    \label{fig:splitting MI temp}
\end{figure}

Using the same interval configuration as in Fig.~\ref{fig:splitting MI} and varying $\beta$, we obtain the dynamics shown in Fig.~\ref{fig:splitting MI temp}. The blue curve is the vacuum result. As expected, it is recovered in the low-temperature limit $\beta\rightarrow\infty$. Increasing the temperature suppresses the mutual information, and above a sufficiently high temperature the mutual information vanishes for all times.

There are two effects of increasing temperature. First, the overall magnitude of the mutual information decreases. Second, some phase transitions disappear. In particular, the transition responsible for the non-analytic peak marked by the black dashed line in the vacuum case is absent already at sufficiently low but finite temperature in the example shown. Equivalently, the system no longer undergoes the corresponding transition in the phases of the individual intervals.

The phase transitions of $A\cup B$ also move closer together as the temperature is increased. Thus the time interval during which the mutual information is non-zero shrinks. At high temperature these transition times approach the quasiparticle light-cone times, although the agreement is never exact in the regime where the mutual information is still non-zero.

\begin{figure}
    \centering
    \includegraphics[width=\linewidth]{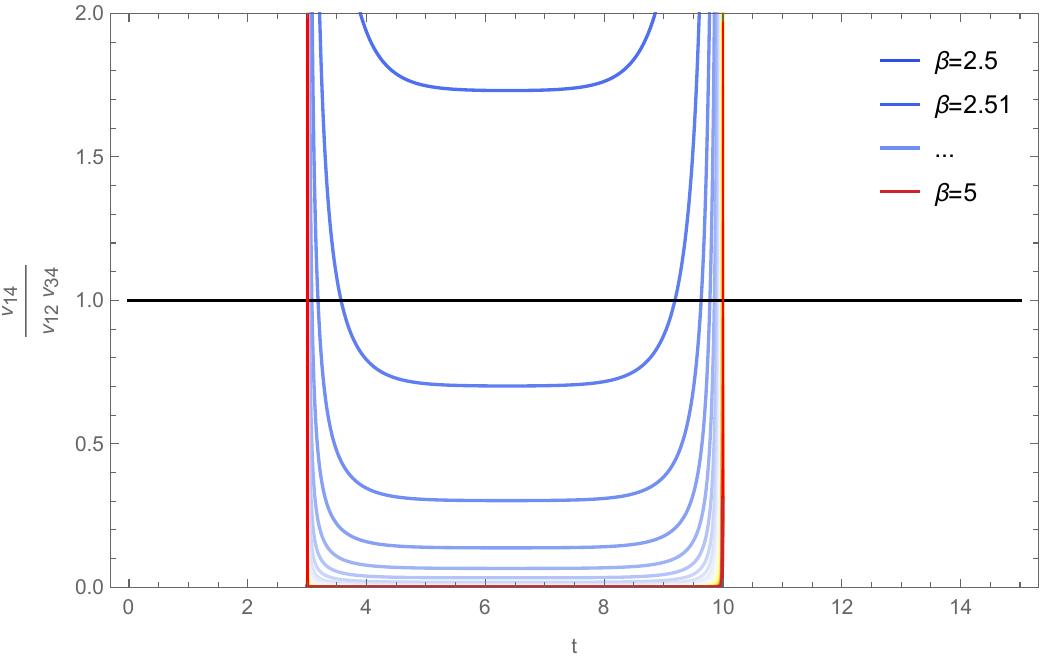}
    \caption{Time dependence of the cross-ratio condition controlling the transition into phase 5a for several inverse temperatures. The horizontal black line is the transition value for $k=1$. At sufficiently high temperature the curve does not cross this value, so the transition does not occur.}
    \label{fig:crossratio over beta}
\end{figure}

This behavior can be understood directly from the cross-ratio conditions. In the vacuum splitting example, the rise of the mutual information is caused by a transition from a decomposable phase to phase 5a. The relevant phase boundary is
\be
\frac{v_{14}(t)}{v_{12}(t)v_{34}(t)}=1,
\ee
for $k=1$. At finite temperature the same condition remains valid, but the functions $v_{ij}(t)$ are modified by the thermal conformal map \eqref{eq:UHPmap-T}. Figure~\ref{fig:crossratio over beta} shows the corresponding time-dependent phase condition for several values of $\beta$. At sufficiently high temperature the curve never crosses the transition value. The system therefore never enters phase 5a, and the mutual-information bump is absent.

Thus the disappearance of mutual information at high temperature is again a statement about the UHP phase structure. Temperature changes the trajectory of the system through the space of cross-ratios. Above a critical temperature, this trajectory never enters any of the non-decomposable phases 4, 5a or 5b, and the mutual information remains zero for all times.

%% file: Sections/5_Finite_c_Effects.tex
Thus far we have focused on holographic conformal field theories, namely CFTs
with large central charge and a sparse spectrum of low-dimension operators. These
assumptions lead to a semiclassical gravitational description and, in particular,
to sharp transitions between competing entanglement saddles. By contrast, many
critical systems realized in condensed-matter or quantum-simulation experiments
are described by CFTs with small central charge, for which the holographic
large-$c$ approximation is not directly applicable. To probe how the picture
developed above is modified away from the holographic limit, we study the
$c=1$ Dirac free fermion CFT using a microscopic lattice regularization. Details
of the lattice implementation, including the choice of Hamiltonian, boundary
conditions, and correlation-matrix methods used to compute entanglement
entropies, are given in Appendix~\ref{app:lattice_details}.

We begin by examining the static phases associated with two intervals on the
upper half-plane. On the lattice, this setup is realized by computing the mutual
information in the ground state of the critical free-fermion chain on a finite
strip with open boundary conditions. The strip is conformally equivalent to the
upper half-plane under the map
\begin{equation}
    z\equiv f(w)=e^{i\pi w/N},
\end{equation}
where $N$ is the length of the strip in lattice units. Pulling back the
upper-half-plane cross-ratios to the strip gives
\begin{equation}
\label{eq:strip_cross_ratio}
    v_{ij}
    =
    \frac{
    \sin^2\!\left[\frac{\pi}{2N}(w_i-w_j)\right]
    }{
    \left|
    \sin\!\left(\frac{\pi w_i}{N}\right)
    \sin\!\left(\frac{\pi w_j}{N}\right)
    \right|
    } .
\end{equation}
Since the configurations considered here are static, we set $w_i=x_i$, with
$x_i$ denoting the position of the corresponding endpoint along the chain.
This allows us to directly compare the lattice data for
$S_A$, $S_B$, $S_{A\cup B}$, and
\begin{equation}
    I_{A:B}=S_A+S_B-S_{A\cup B}
\end{equation}
to the corresponding holographic predictions expressed in terms of the same
cross-ratios. In the lattice calculation we choose free open ends, which have
vanishing boundary entropy in the continuum limit. We therefore set the
boundary entropy parameter to $k=1$ in the holographic comparison  (as $\ln(1)=0$).

\begin{figure*}
    \centering
    \includegraphics[width=\linewidth]{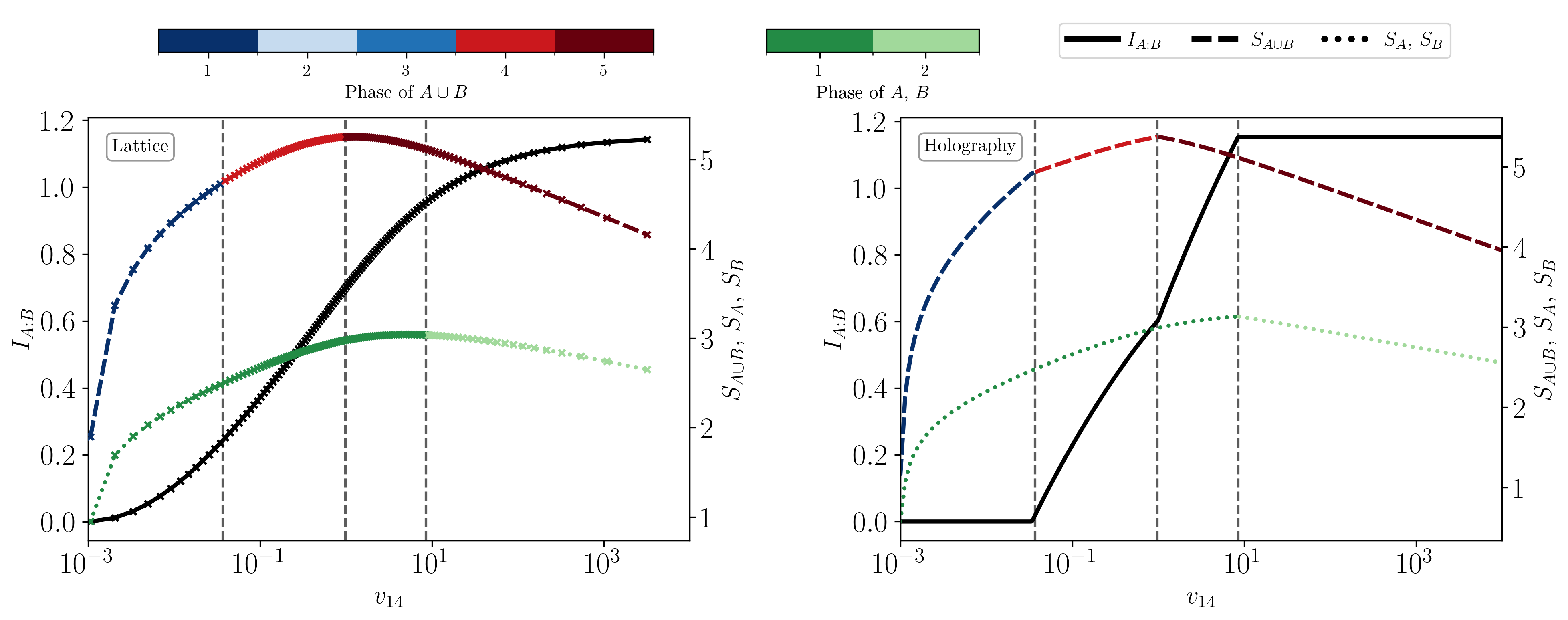}
    \caption{
    Static mutual information and entanglement entropies for two intervals on a
    strip. The left panel shows the free-fermion lattice result for a chain with
    $N=4000$ sites and open boundary conditions. The right panel shows the
    corresponding holographic calculation with a boundary. We plot
    $I_{A:B}$, $S_{A\cup B}$, $S_A$, and $S_B$ as functions of the
    cross-ratio $v_{14}$ defined in Eq.~\eqref{eq:strip_cross_ratio}. The
    intervals $A$ and $B$ have fixed length, here $80$ lattice sites, and are
    arranged symmetrically about the center of the chain. Increasing their
    separation increases $v_{14}$. The sharp holographic phase transitions are
    rounded in the $c=1$ free-fermion theory.
    }
    \label{fig:Lattice_Holo_static_phases}
\end{figure*}

The result is shown in Fig.~\ref{fig:Lattice_Holo_static_phases}. The left
panel displays the free-fermion lattice data for a chain with $N=4000$ sites and
open boundary conditions, while the right panel shows the corresponding
holographic calculation with boundaries. The intervals $A$ and $B$ are taken to
have fixed length and are placed symmetrically about the center of the chain.
Their separation is then varied, which changes the relevant cross-ratios. In the
holographic result, the entropies are controlled by a small number of competing
geodesic configurations, leading to sharp transitions as the dominant saddle
changes. In the $c=1$ free-fermion theory these sharp transitions are smoothed
out. This is consistent with the expectation that, away from the large-$c$
sparse-spectrum regime, additional conformal blocks contribute significantly to
the relevant correlation functions, thereby washing out the non-analyticities that
appear in the holographic approximation, as discussed in
Sec.~\ref{sec:section2}.

\begin{figure}
    \centering
    \includegraphics[width=\linewidth]{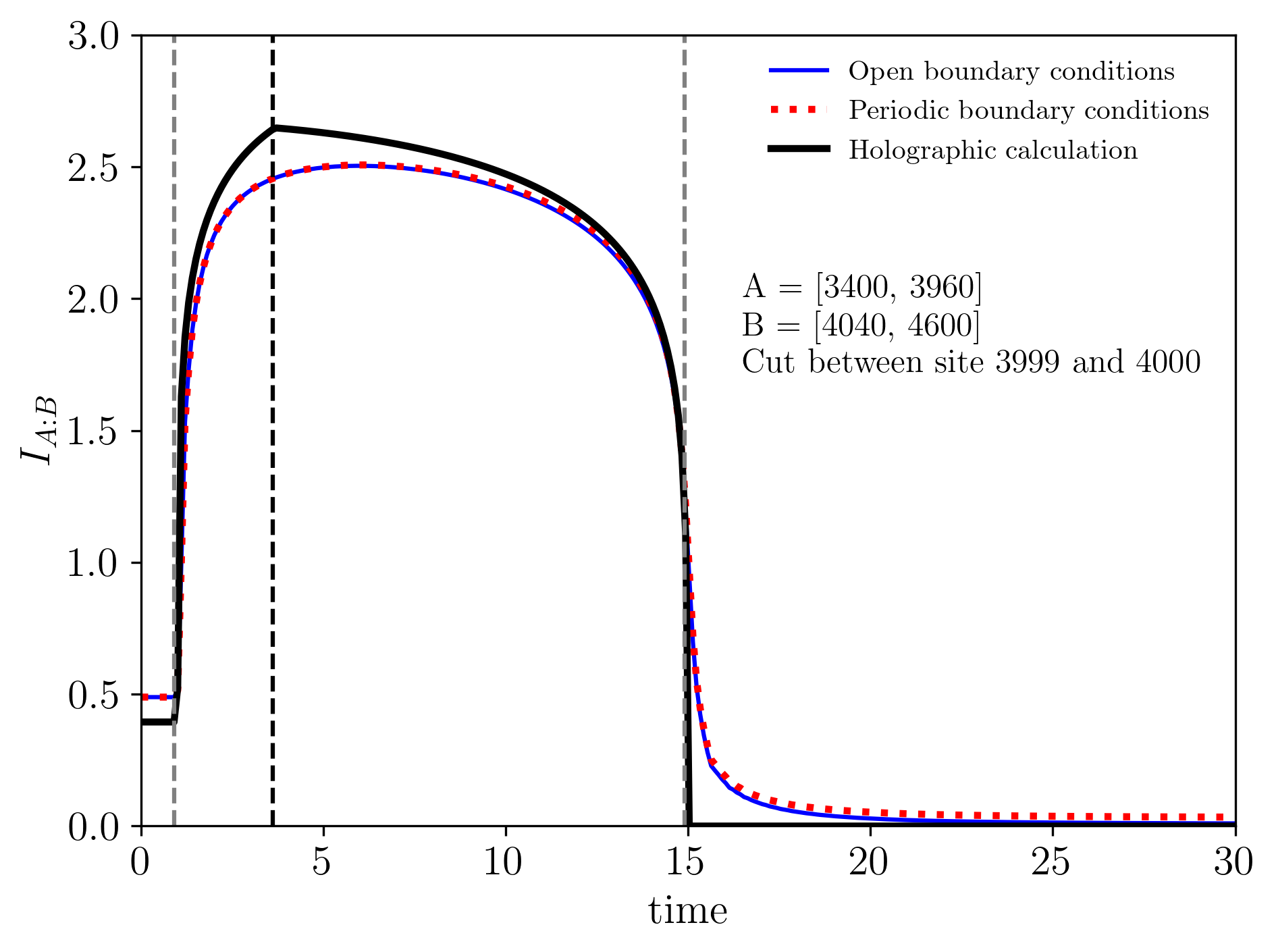}
    \caption{
    Time evolution of the mutual information after the splitting quench for the
    symmetric interval configuration corresponding to
    Fig.~\ref{fig:splitting MI peak}. We compare free-fermion lattice results
    with open and periodic boundary conditions to the holographic prediction.
    The lattice has $N=8000$, $\gamma=20$, and $a=1/40$, so that
    $2\gamma a=1$. The chain is split at $t=0$ between sites $3999$ and
    $4000$. The agreement between open and periodic boundary conditions
    indicates that boundary reflections are negligible on the time scales shown.
    }
    \label{fig:Lattice_splitting_quench_sym}
\end{figure}

We now turn to the real-time splitting quench, which is the main focus of this
work. Figure~\ref{fig:Lattice_splitting_quench_sym} shows the lattice analogue
of the symmetric configuration considered in Fig.~\ref{fig:splitting MI}: the
two subsystems are placed symmetrically with respect to the splitting point. The
initial state is the ground state of the connected chain. At $t=0$ the hopping
term across the midpoint is removed, and the state is evolved with the
Hamiltonian of the split system. We compare open and periodic boundary
conditions for a large chain with $N=8000$, $\gamma=20$, and $a=1/40$, so that
the low-energy propagation velocity is expected to be
\begin{equation}
    v_g \simeq 2\gamma a = 1 .
\end{equation}
In Appendix~\ref{app:lattice_details}, we demonstrate numerically in a system with $4000$ sites that the dominant velocity of the main excitation after the splitting quench is $v=0.989882\pm 0.000095$.
The agreement between open and periodic boundary conditions shows that, for the
times and subsystem sizes considered, effects from the physical boundaries
are negligible. Therefore, we expect that boundary effects do not contribute and we can reasonably compare the lattice data to the
holographic result on the infinite line.

In the holographic calculation the time dependence of the mutual information
contains three sharp transitions. The first is the onset of mutual information,
occurring at approximately $t\sim 1$. The second is a transition between two
distinct nonzero-mutual-information phases, occurring at approximately
$t\sim 4$. The third is the disappearance of mutual information, occurring at
approximately $t\sim 15$. The lattice data allow us to ask which aspects of this
structure survive at small central charge.

\begin{figure}
    \centering
    \includegraphics[width=\linewidth]{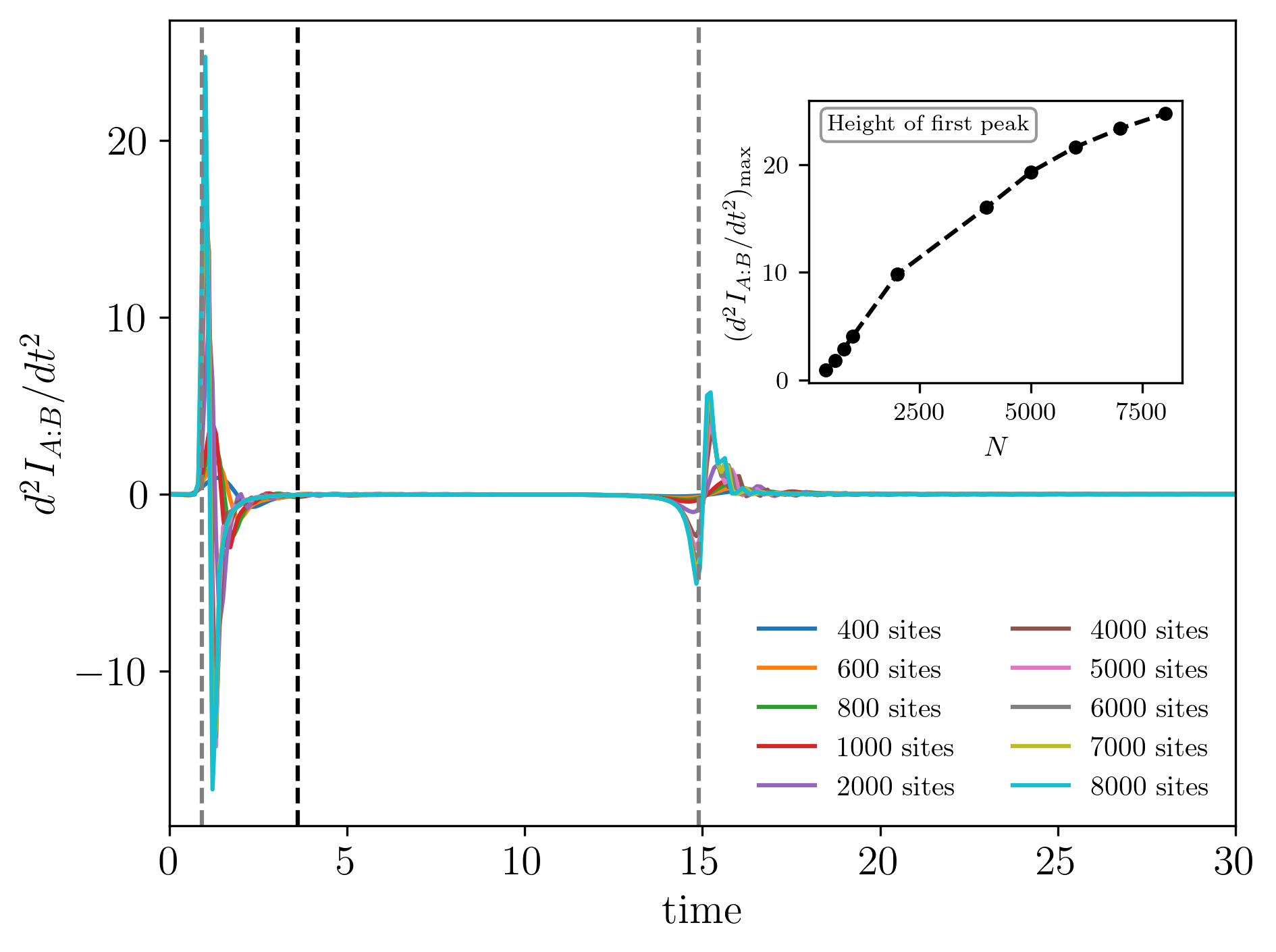}
    \caption{
    Second time derivative of the mutual information during the splitting quench
    shown in Fig.~\ref{fig:Lattice_splitting_quench_sym}. We compare several
    lattice sizes $N$ while keeping the physical size $Na$ and the low-energy
    velocity $|v_g|\simeq 2\gamma a$ fixed. Increasing $N$ therefore corresponds
    to decreasing the lattice spacing and approaching the continuum limit. The
    inset shows the maximum of the first peak of
    $d^2 I_{A:B}/dt^2$ as a function of system size. The vertical dashed lines
    indicate the locations of the phase transitions predicted by the
    holographic calculation.
    }
    \label{fig:Lattice_2nd_derivative_scaling}
\end{figure}

The lattice result does not treat every transition equally. The intermediate transition between
two nonzero-mutual-information phases is smoothed out in the free-fermion
calculation. This is again consistent with the expectations of Sec. \ref{sec:section2} for $c=1$. In contrast, the sharp transitions associated with the appearance and
disappearance of mutual information remain. The smoothing visible in the plot is a finite size effect\footnote{Note that because CFTs have no length scale, there is no distinction between finite size effects (long wavelength) and discretization effects (short wavelength)}. We study the scaling of the second time derivative of the mutual
information. The result is shown in
Fig.~\ref{fig:Lattice_2nd_derivative_scaling}. We vary the number of lattice
sites $N$ while keeping both the physical system size $Na$ and the low-energy
velocity $2\gamma a$ fixed. Thus increasing $N$ corresponds to decreasing the
lattice spacing and approaching the continuum limit. The peaks in
$d^2 I_{A:B}/dt^2$ associated with the onset and disappearance of mutual
information become sharper as $N$ is increased. The inset shows the growth of
the first peak with system size, indicating that the curvature at the transition
diverges in the continuum limit – indicating non-analytic behavior.
Taken together, the free-fermion results support the interpretation that
the symmetry-controlled transitions governing the presence or absence of mutual
information are not special to holographic CFTs. Rather, they appear to persist in non-holographic conformal field theories, where the transitions between different phases of mutual information get smoothed out by $1/c$ corrections.

%% file: Sections/6_Discussion.tex
In this paper we analyzed the dynamical phase structure of
mutual information in $1+1$-dimensional conformal field theory following local splitting (and joining) quenches.
Working in the holographic limit – large central charge and sparse spectrum of low-dimensional operators – we obtained the following results.

First, we classified all possible phases of the entanglement entropy of two disjoint intervals, $S_{X \cup Y}$, on the upper half-plane.
Of the naively possible pairings of four boundary-anchored geodesics, exactly seven contribute;
these are listed in eq.~\eqref{eq:AuB phases} and depicted in Fig.~\ref{fig:Phases}.
Four of those seven phases satisfy $S_{X\cup Y} = S_X + S_Y$ and therefore carry no mutual
information; the remaining three phases do carry mutual information.
Both collections of phases are invariant under a distinct $D_4$ subgroup of the permutation group
acting on the interval endpoints. The conditions for transitioning from one phase to another can be expressed as conformally invariant cross-ratio inequalities
defined in eq.~\eqref{eq:crossratio} and tabulated in Appendix~\ref{app:phases}.

Second, we demonstrated that the world-sheets we consider are conformally equivalent
to the upper half-plane.
This means the entire static phase structure can be pulled back to Lorentzian time via a single
conformal map specific to each protocol, turning the cross-ratio conditions into explicit
time-dependent inequalities.
We carried out this program explicitly for the splitting quench and the joining quench, both
in the vacuum and at finite temperature, and found that the mutual information generically
undergoes multiple dynamical phase transitions.
Two qualitatively distinct types of transition arise.
The first type is associated with a change in the phase of the composite system $A\cup B$ and governs whether there is or is not  mutual information.
The second type is associated with change in the phases of the individual
subsystems $A$ and $B$ and produces transitions between different phases of mutual information.
In both cases the transition times are fully determined by when the appropriate cross-ratio
condition crosses its threshold, providing an alternative to the
quasiparticle approximation which misses non-analytic behavior arising from transition between mutual-information carrying phases.

Third, we studied finite central charge effects by simulating the $c = 1$ Dirac free fermion theory on
a lattice. In particular, we investigated both the ground state boundary phase structure and dynamical phases after a local splitting quench.
The sharp holographic transitions between distinct nonzero-mutual-information phases are smoothed
out at finite $c$, consistent with the contribution of additional conformal blocks away from the
large-$c$ sparse-spectrum regime.
By contrast, the transitions associated with the onset and disappearance of mutual information
are robust: the second time derivative of the mutual information increases systematically as the
continuum limit is approached, suggesting that these transitions are governed by the $D_4$
symmetry of the cross-ratios and
are not an artifact of the large-$c$ limit. Let us stress that any realistic experiment can only access theories with central charges of order one. It is therefore important that the qualitative behavior observed at finite central charge remains close to the holographic result. In particular, the similarity between the two plots in Fig.~15, corresponding to $c=1$ and $c=\infty$, suggests only a weak dependence on $c$. This provides encouraging evidence that the phenomena identified in the holographic limit may have realistic prospects for experimental observation.

The results reported here connect three strands of the literature in a new way. The dynamical quantum phase transition literature, initiated in~\cite{heyl2013} and reviewed
in~\cite{heyl2018}, has established that non-analytic behavior can arise in real-time evolution.
Our work provides a concrete, analytically tractable realization within $1+1$-dimensional CFT:
the mutual information of two intervals following a local quench undergoes a sequence of such
transitions, and the transitions are governed by the crossing of conformally invariant
cross-ratio thresholds rather than by light-cone arrivals. This makes the mechanism transparent and opens a systematic route to computing finite-$c$ and
finite-temperature corrections.
The quasiparticle picture~\cite{calabrese2005c,calabrese2009a} correctly identifies the gross
causal structure of post-quench entanglement dynamics, but it cannot account for the
non-analytic peaks and extended tails we observe.
Our framework supersedes the quasiparticle description for the mutual information: the precise
transition times, and the distinction between phases that do and do not carry mutual
information, follow directly from the static UHP phase structure.
At finite temperature the phase transitions associated with mutual information creation and
annihilation converge toward the quasiparticle predictions only in the high-temperature limit,
where the relevant cross-ratio conditions fail to cross their thresholds and the mutual
information is suppressed entirely, in agreement with the finite-temperature results of~\cite{fischler2013}.

Finally, our identification of the $D_4$ symmetry governing the presence or absence of mutual
information suggests a Landau-type organizing principle for non-equilibrium entanglement dynamics.
In equilibrium, Landau theory associates each phase with the symmetry of its order parameter.
Here we find an analogous structure in real time: the phases of $A\cup B$ that carry mutual
information are those invariant under one $D_4$ subgroup of endpoint permutations, while those
that do not carry mutual information are invariant under a different $D_4$ subgroup.
The dynamical onset of mutual information is therefore heralded by the breaking of the latter symmetry in
favor of the former, with the associated non-analyticity playing the role of a phase boundary
in time. This symmetry is genuinely dynamical: it does not correspond to the breaking of an equilibrium
symmetry~\cite{heyl2014}, but to a symmetry structure that exists only in the real-time
evolution of mutual information.

The most immediate open question raised by our results is whether the Landau analogy can be deepened. In equilibrium, one can construct a Landau effective action purely from the symmetry and the order parameter, without reference to the microscopic theory, and use it to classify
universality classes and compute critical exponents.
Here the $D_4$ subgroups of the endpoint permutation group play the role of the symmetry, and
the mutual information itself is a candidate order parameter.
A natural question is whether an effective action can be written whose saddle points reproduce
the phase structure we found, and whether the non-analytic behavior at the transition times can
be captured by such an action in a universal way.
A related question is whether the non-analyticity in the mutual information implies non-analytic
behavior in the Loschmidt amplitude at the same transition times.
In the Lee-Yang picture of dynamical quantum phase transitions, non-analytic behavior
in the return amplitude is associated with the crossing of zeros in the complex time plane.
Whether the dynamical phase transitions we identify here are accompanied by such zeros---and
whether the Loschmidt amplitude can be expressed as a function of the mutual information ---is an interesting avenue to explore.

A second direction is to investigate how the phase structure changes when the large-$c$
assumption is relaxed in a controlled way.
The compact free boson at radius $R$ is a minimal example: it has $c = 1$ and therefore lies
entirely outside the holographic regime, yet its partition function on a genus-$g$ surface is
known in closed form~\cite{alvarez-gaume1987,Calabrese_2009,Calabrese_2011}.
This would allow an explicit computation of the replica partition function on the relevant
branched-cover geometry without invoking the holographic approximation.
Because the torus partition function and the R\'{e}nyi-2 entropy of a single interval are
analytic in the modular parameter for the compact free boson~\cite{headrick2010a}, we expect the non-analytic
behavior in the mutual information to arise from the $n\rightarrow 1$ limit.
Studying this model would therefore sharply test whether the $D_4$-symmetry-controlled
transitions survive away from the large-$c$ limit even when the intermediate transitions do not,
complementing the numerical evidence from the free fermion presented in Sec.~\ref{sec:section5}.

A third direction concerns the extension to more complicated quench geometries, such as multiple splitting quenches~\cite{lap2025a}. From a holographic perspective, each additional split introduces a new end-of-the-world brane, and the minimal geodesic connecting a pair of interval endpoints can now terminate on any of
these branes. As such we expect a much richer phase structure.

%% file: Sections/Appendix_A_Conformal_Maps.tex
In the main text, we use conformal maps to uniformize the various world-sheets to the upper half-plane.\\

The Riemann mapping theorem states that all simply-connected subsets of the complex plane are conformally equivalent to the upper half-plane~\cite{riemann2013}. This is an existence statement, and the construction of the appropriate maps is more subtle (a classic reference is~\cite{nehari1952}).\\

For convenience, we summarize the elementary maps used throughout the main text:\\
Splitting quench on the line:
\begin{equation}
    z = i\sqrt{\frac{w+ia}{w-ia}}
\end{equation}
Joining quench on the line:
\begin{equation}
    z = i\sqrt{\frac{ia-w}{ia+w}}
\end{equation}
Splitting quench on the line at finite temperature:
\begin{equation}
    z = i\sqrt{\frac{\tanh(\frac{\pi}{\beta} w)+\tanh(\frac{i\pi}{\beta} a)}{\tanh(\frac{\pi}{\beta} w)-\tanh(\frac{i\pi}{\beta} a)}}
\end{equation}

Joining quench on the line at finite temperature:
\begin{equation}
    z = i\sqrt{\frac{\tanh(\frac{i\pi}{\beta} a)-\tanh(\frac{\pi}{\beta} w)}{\tanh(\frac{i\pi}{\beta} a)+\tanh(\frac{\pi}{\beta} w)}}
\end{equation}
Splitting quench on the circle:
\begin{equation}
    z = i\sqrt{\frac{e^{\frac{2\pi}{L} w}+e^{\frac{2\pi i}{L} a}}{e^{\frac{2\pi}{L} w}+e^{-\frac{2\pi i}{L} a}}}
\end{equation}
Joining quench on the circle:
\begin{equation}
    z = i\sqrt{\frac{e^{\frac{2\pi i}{L} a}-e^{\frac{2\pi }{L} w}}{e^{\frac{2\pi i}{L} a}+e^{\frac{2\pi }{L} w}}}
\end{equation}
These maps are not unique in mapping to the UHP, but are convenient, since they reduce to the simple vacuum splitting and joining maps in the limits $\beta \rightarrow \infty$ and $L \rightarrow \infty$, respectively.

%% file: Sections/Appendix_B_Phase_Transition_Inequalities.tex
\label{app:table of phase conditions}
The procedure described in Sec.~\ref{sec:section3} can be used to derive all phase transition points, which leads to the conditions given in Tab.~\ref{tab:phase conditions} 
\begin{table}
    \centering
    \resizebox{\columnwidth}{!}{%
    \begin{tabular}{| c | c | c | c | c | c | c |}
    \hline
        & & & & & & \\    
        1 & & & & & & \\
        & & & & & & \\    
    \hline
        & & & & & & \\
        $v_{12}v_{34} < k^2$ & 2 & & & & & \\
        & & & & & & \\
    \hline
        & & & & & & \\
        $v_{34} < k$ & $\frac{1}{v_{12}} < \frac{1}{k}$ & 3a & & & & \\
        & & & & & & \\
    \hline
        & & & & & & \\
        $v_{12} < k$ & $\frac{1}{v_{34}} < \frac{1}{k}$ & $\frac{v_{12}}{v_{34}} < 1$ & 3b & & & \\
        & & & & & & \\
    \hline
        & & & & & & \\
        $\frac{v_{12}v_{34}}{v_{14}v_{23}} < 1$ & $\frac{1}{v_{23}v_{14}} < \frac{1}{k^2}$ & $\frac{v_{12}}{v_{23}v_{14}} < \frac{1}{k}$ & $\frac{v_{34}}{v_{23}v_{14}} < \frac{1}{k}$ & 4 & & \\
        & & & & & & \\
    \hline
        & & & & & & \\
        $\frac{v_{12}v_{34}}{v_{14}} < k$ & $\frac{1}{v_{14}} < \frac{1}{k}$ & $\frac{v_{12}}{v_{14}} < 1$ & $\frac{v_{34}}{v_{14}} < 1$ & $v_{23} < k$ & 5a & \\
        & & & & & & \\
    \hline
        & & & & & & \\
        $\frac{v_{12}v_{34}}{v_{23}} < k$ & $\frac{1}{v_{23}} < \frac{1}{k}$ & $\frac{v_{12}}{v_{23}} < 1$ & $\frac{v_{34}}{v_{23}} < 1$ & $v_{14} < k$ & $\frac{v_{14}}{v_{23}} < 1$ & ~5b~ \\
        & & & & & & \\
    \hline
    \end{tabular}
    }
\caption{Cross-ratio conditions for the subsystem $X \cup Y$ to be in a certain phase. Each phase is defined by six conditions listed in the rows and columns intersecting the phase number. The conditions in the column for each number need to be met as stated; the conditions in the row of each phase number must be negated, i.e., the inequality must be reversed.}
\label{tab:phase conditions}
\end{table}

%% file: Sections/Appendix_C_Group_Structure_of_Phases.tex
On the upper half-plane, the mutual information \(I(X:Y)\) may be regarded as a function of the four endpoint coordinates, $X=[z_1,z_2]$ and $Y=[z_3,z_4]$:

\[
I(X:Y)=I(z_1,z_2,z_3,z_4), \qquad z_i\in\mathbb{H}.
\]
Since the calculation involves a conformal four-point function of identical operators, one might initially expect the result to be invariant under the full permutation group \(S_4\).\footnote{A similar line of reasoning was used in~\cite{dymarsky2018} to derive nontrivial crossing constraints on stress-tensor four-point functions.}  However, the phases of $S_{X\cup Y}$ relevant for mutual information are not invariant under all of $S_4$. Once one distinguishes between phases with vanishing and non-vanishing mutual information, the relevant symmetry is reduced. Both phases are invariant under different $D_4$ subgroups of $S_4$.\\

 Consider Fig.~\ref{fig:Phases} of all possible phases of $S_{X\cup Y}$. The phases with non-vanishing mutual information are invariant under the exchange \(z_1\leftrightarrow z_4\), whereas the phases with vanishing mutual information are not. Conversely, the phases with vanishing mutual information are invariant under \(z_1\leftrightarrow z_2\), whereas the phases with non-vanishing mutual information are not. The relevant question is therefore: which subgroup of \(S_4\) leaves each class of phases invariant?

The phases with vanishing mutual information are invariant under the subgroup that leaves $X$ and $Y$ invariant:
\[
\mathrm{Stab}(\{\{1,2\},\{3,4\}\})\simeq D_4,
\]
while the phases with non-vanishing mutual information are invariant under\footnote{Where Stab is the stabilizer subgroup of this pairing. This is analogous to Wigner's classification of particle states by noting which subgroup of the Poincar\'e group massless versus massive states are invariant under~\cite{wigner1989}.}
\[
\mathrm{Stab}(\{\{1,4\},\{2,3\}\})\simeq D_4.
\]
Here \(D_4\) denotes the symmetry group of the square, of order \(8\).

Importantly, although both classes are invariant under a subgroup isomorphic to \(D_4\), these are distinct \(D_4\) subgroups of \(S_4\). Their intersection is
\[
\{e,(12)(34),(13)(24),(14)(23)\}\simeq V_4,
\]
where \(V_4\simeq \mathbb{Z}_2\times \mathbb{Z}_2\) is the Klein four-group. This common subgroup is precisely the group of ``kinematic permutations" discussed in~\cite{dymarsky2018,kravchuk2018}; equivalently, it is the subgroup of permutations that leaves the cross-ratio
\[
u=\frac{|z_{12}||z_{34}|}{|z_{13}||z_{24}|}
\]
invariant.

In summary, the full permutation symmetry of four identical operator insertions is \(S_4\), which has \(24\) elements. The phases with vanishing and non-vanishing mutual information are each invariant under a distinct \(D_4\) subgroup, each with \(8\) elements, and these two subgroups intersect in the kinematic permutation group \(V_4\), which has \(4\) elements. This suggests the following symmetry-breaking pattern: in a dynamical process where mutual information becomes nonzero at time \(t_C\), the \(D_4\) symmetry associated with the phase of vanishing mutual information is reduced to the common \(V_4\) subgroup at \(t_C\). However at $t_C$ the $V_4$ subgroup is also restored to the other $D_4$ group\footnote{There is another $D_4$ subgroup of $S_4$, but it is associated with the group of phases which do never contribute because the geodesics are always longer (See Sec.~\ref{sec:section3}).}. This suggests a non-trivial symmetry breaking pattern that merits further study.

\begin{figure}
    \centering
    \includegraphics[width=.8\linewidth]{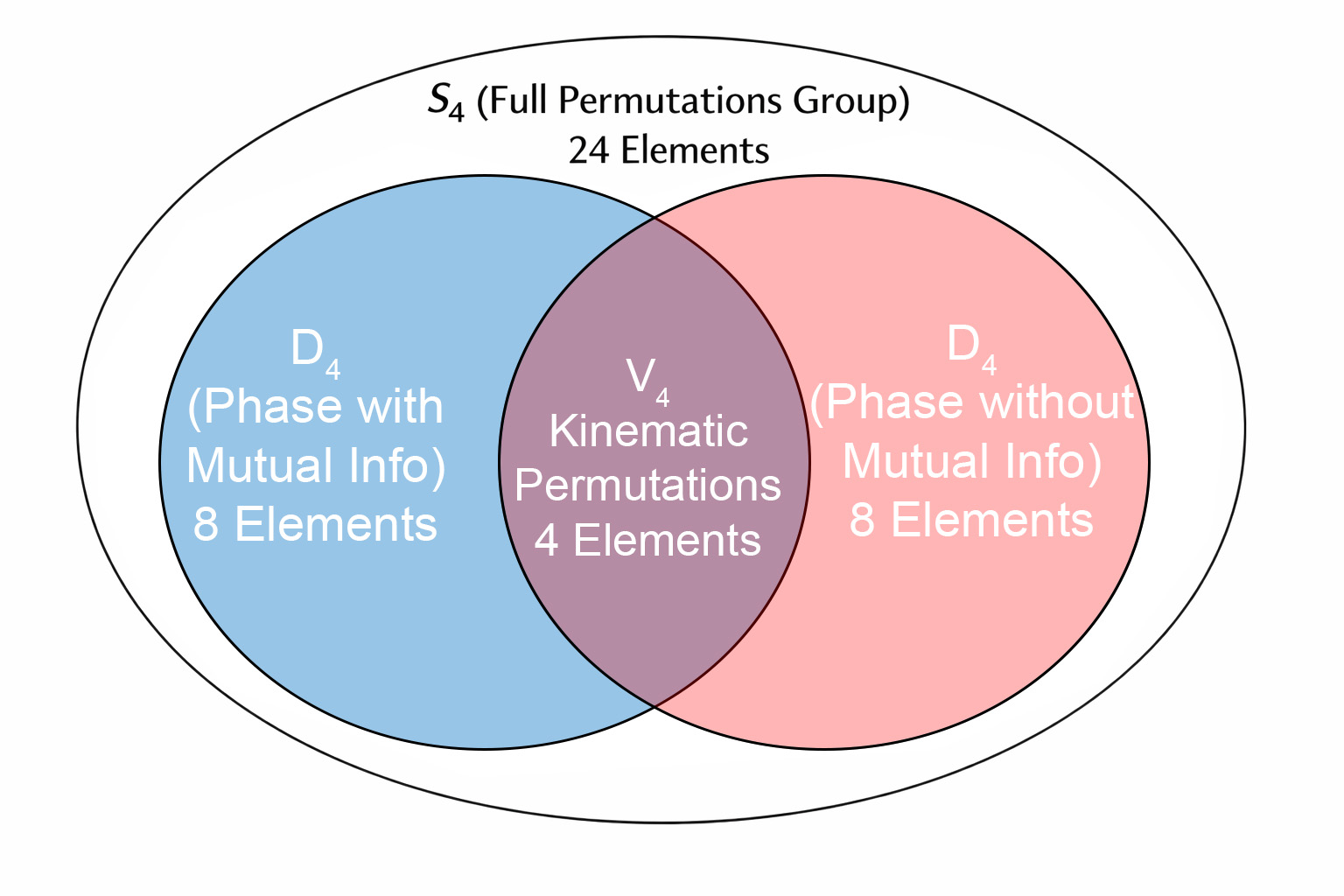}
    \caption{Distinct $D_4$ subgroups of $S_4$ associated with phases of $S_{X\cup Y}$ with vanishing and non-vanishing mutual information, and their common $V_4$ subgroup.}
    \label{fig:groupstructure}
\end{figure}

%% file: Sections/Appendix_D_Lattice_Details.tex
In this Appendix we describe the lattice regularization used to simulate the
$1+1$-dimensional free Dirac conformal field theory with central charge $c=1$.
The continuum theory arises as the low-energy, long-wavelength limit of the
one-dimensional nearest-neighbor hopping Hamiltonian
\begin{equation}
\label{eq:Free_fermion_Ham}
    H=-\gamma \sum_{i=i_1}^{i_N}
    \left(c_i^{\dagger} c_{i+1}+c^{\dagger}_{i+1} c_i\right),
\end{equation}
where $c_i$ and $c_i^\dagger$ are spinless fermionic annihilation and creation
operators on lattice site $i$. The lattice spacing is denoted by $a$, and the
canonical anticommutation relations are
\begin{equation}
    \{c_i,c_j^\dagger\}=\delta_{ij}, 
    \qquad 
    \{c_i,c_j\}=\{c_i^\dagger,c_j^\dagger\}=0 .
\end{equation}
We work at half filling, where the Fermi points lie at
$k_F=\pm \pi/(2a)$. Linearizing the spectrum about these two Fermi points gives
a relativistic Dirac fermion in the continuum limit, with central charge
$c=1$~\cite{affleck1990,ginsparg1991,difrancesco1997}.

We consider both periodic and open boundary conditions in order to verify that
our continuum results are insensitive to the choice of lattice boundary
conditions. For periodic boundary conditions (PBC), we take sites
$i=0,\ldots,N-1$ and identify
\begin{equation}
    c_N \equiv c_0 .
\end{equation}
In this case the sum in Eq.~\eqref{eq:Free_fermion_Ham} runs from
$i_1=0$ to $i_N=N-1$. For open boundary conditions (OBC), the sites again run
from $0$ to $N-1$, but the hopping term connecting the two ends is absent; the
sum then runs from $i_1=0$ to $i_N=N-2$.

For PBC, the Hamiltonian is diagonalized by the Fourier transform
\begin{equation}
    c_j=\frac{1}{\sqrt{N}}\sum_k e^{ikaj}\,\tilde c_k ,
    \qquad 
    k=\frac{2\pi n}{Na}, 
    \qquad n=0,\ldots,N-1 .
\end{equation}
The Hamiltonian becomes
\begin{equation}
    H=\sum_k \epsilon_k \tilde c_k^\dagger \tilde c_k ,
    \qquad 
    \epsilon_k=-2\gamma \cos(ka).
\end{equation}
At half filling, the many-body ground state is obtained by filling all
single-particle modes with negative energy. Equivalently, the occupied modes
satisfy
\begin{equation}
    -\frac{\pi}{2a}<k<\frac{\pi}{2a},
\end{equation}
up to the usual finite-size convention for modes exactly at the Fermi points.
The low-energy excitations are particle-hole excitations near
$k=\pm k_F=\pm \pi/(2a)$. Expanding the dispersion relation around the two
Fermi points gives
\begin{equation}
    \epsilon_{k_F+q}
    =
    -2\gamma \cos\left(\frac{\pi}{2}+qa\right)
    \simeq 2\gamma a\, q ,
\end{equation}
and
\begin{equation}
    \epsilon_{-k_F+q}
    =
    -2\gamma \cos\left(-\frac{\pi}{2}+qa\right)
    \simeq -2\gamma a\, q ,
\end{equation}
for $|qa|\ll 1$. Thus the continuum limit contains right- and left-moving
relativistic fermions with velocity
\begin{equation}
    v_F = 2\gamma a .
\end{equation}
This Fermi velocity is the maximal velocity of the low-energy continuum theory.
In the lattice model the exact group velocity is
\begin{equation}
    v(k)=\frac{\partial \epsilon_k}{\partial k}
    =2\gamma a \sin(ka),
\end{equation}
whose maximum magnitude is also $2\gamma a$. Throughout this work we set the
continuum speed of light to one by choosing
\begin{equation}
    2\gamma a=1 .
\end{equation}

For OBC, the single-particle eigenmodes are standing waves rather than plane
waves. The open chain is described at low energies by a boundary conformal
field theory. We consider free lattice ends, which correspond in the bosonized
continuum description to Dirichlet boundary conditions~\cite{eggert1992}. For
this conformal boundary condition the Affleck-Ludwig boundary entropy is
\begin{equation}
    S_{\rm bdy}=\ln g,
    \qquad g=1 ,
\end{equation}
so that $S_{\rm bdy}=0$~\cite{fagotti2011}. This absence of an additional
constant boundary contribution is useful when comparing lattice entanglement
data with continuum CFT predictions.

We compute ground-state entanglement entropies using the fact that the ground
state of Eq.~\eqref{eq:Free_fermion_Ham} is Gaussian. Therefore all reduced
density matrices are completely determined by the two-point correlation
function. For a subsystem $A$ consisting of $l$ lattice sites, we define the
restricted correlation matrix
\begin{equation}
    C_{ij}
    =
    \langle \psi_0 | c_i^\dagger c_j | \psi_0\rangle,
    \qquad i,j\in A ,
\end{equation}
where $|\psi_0\rangle$ is the many-body ground state. The reduced density
matrix has the Gaussian form
\begin{equation}
    \rho_A = \frac{e^{-\mathcal H_A}}{Z_A},
    \qquad
    \mathcal H_A=\sum_{i,j\in A} h_{ij} c_i^\dagger c_j ,
\end{equation}
where the single-particle entanglement Hamiltonian is related to the correlation
matrix by
\begin{equation}
    h=\ln\left(\frac{1-C}{C}\right).
\end{equation}
Let $\nu_\alpha$, $\alpha=1,\ldots,l$, denote the eigenvalues of $C$. The von
Neumann entropy of $A$ is then
\begin{equation}
\label{eq:entropy_corr_matrix}
    S_A
    =
    -\sum_{\alpha=1}^{l}
    \left[
        \nu_\alpha \ln \nu_\alpha
        +
        (1-\nu_\alpha)\ln(1-\nu_\alpha)
    \right].
\end{equation}
This method is numerically efficient because it requires diagonalizing only the
$l\times l$ correlation matrix restricted to the subsystem rather than the full
many-body density matrix~\cite{peschel2003,chung2001}.

We now describe the real-time splitting protocol used in the main text. We
first prepare the ground state $|\psi_0\rangle$ of the connected Hamiltonian
$H$ in Eq.~\eqref{eq:Free_fermion_Ham}. At time $t=0$ we cut the chain between
sites $i'$ and $i'+1$ by removing the hopping term connecting these two sites.
Equivalently, the post-quench Hamiltonian is
\begin{equation}
\label{eq:split_hamiltonian}
    H'
    =
    H+\gamma
    \left(
        c_{i'}^\dagger c_{i'+1}
        +
        c_{i'+1}^\dagger c_{i'}
    \right),
\end{equation}
which cancels the corresponding hopping term in $H$. The state is then evolved
with the split Hamiltonian,
\begin{equation}
    |\psi(t)\rangle = e^{-iH't}|\psi_0\rangle .
\end{equation}
For OBC, this operation separates the original chain into two disconnected
subchains, with Hamiltonians $H_1$ and $H_2$. For PBC, cutting one bond turns
the ring into an open chain and introduces two free boundaries at the cut. In
both cases we choose the cut to lie at the center of the system,
\begin{equation}
    i'=\frac{N}{2}-1 ,
\end{equation}
for even $N$. The total system size is chosen sufficiently large that excitations
created by the quench do not reach the physical ends of the system during the
time interval of interest. Thus boundary reflections do not affect the data
shown in the main text; see Fig.~\ref{fig:Lattice_splitting_quench_sym}.

The splitting quench injects energy locally near the cut and produces
ballistically propagating wave packets. In the continuum CFT there is a single
propagation velocity, which in our units is $v=1$. On the lattice, however, the
dispersion relation is nonlinear away from the Fermi points, and therefore
lattice excitations can propagate with a range of group velocities
\begin{equation}
    v(k)=2\gamma a\sin(ka).
\end{equation}
The universal low-energy contribution is controlled by the modes near the Fermi
points and propagates with velocity $v_F=2\gamma a$. Consequently, after setting
$2\gamma a=1$, the dominant low-energy signal is expected to move at unit
velocity.

\begin{figure}[H]
    \centering
    \includegraphics[width=\linewidth]{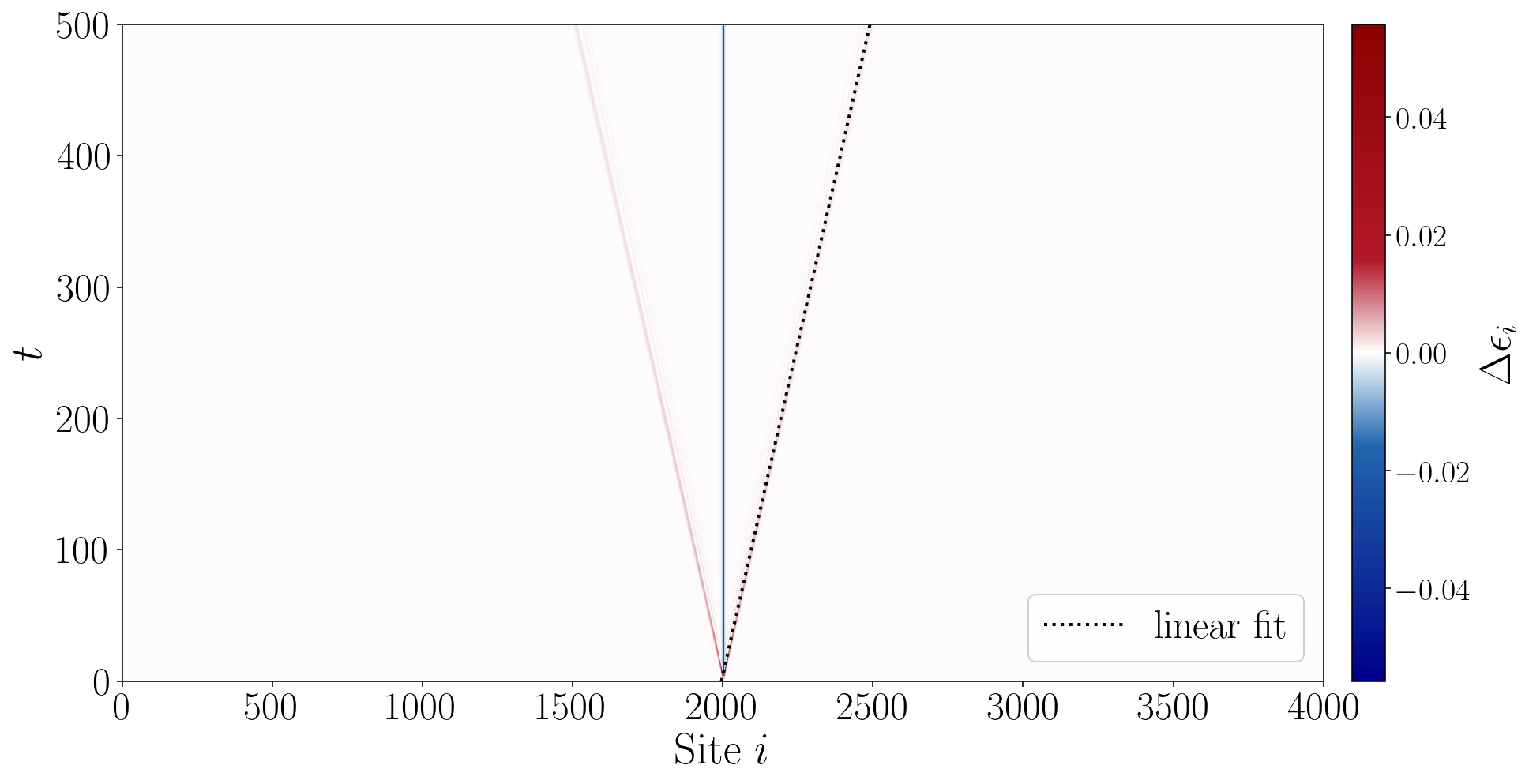}
    \caption{
    Time evolution of the excess local energy density $\Delta \epsilon_i$,
    obtained by subtracting the static ground-state energy density of the
    uncut chain. The chain has $N=4000$, $\gamma=0.5$, and $a=1$, so that
    the continuum velocity is $2\gamma a=1$. The chain is split at $t=0$
    between sites $1999$ and $2000$. A linear fit $x(t)=vt+b$ to the
    right-moving energy-density maximum gives
    $v=0.989882\pm 0.000095$.
    }
    \label{fig:Lattice_heatmap}
\end{figure}

This expectation is confirmed numerically in Fig.~\ref{fig:Lattice_heatmap}.
There we show the time evolution of the excess local energy density
$\Delta \epsilon_i$, defined as the local energy density after the split with
the static ground-state energy density of the uncut chain subtracted. The data
are for a chain with $N=4000$ sites, $\gamma=0.5$, and $a=1$, so that
$2\gamma a=1$. The cut is performed at $t=0$ between sites $1999$ and $2000$.
Fitting the trajectory of the maximum of the right-moving energy-density packet
to a linear form $x(t)=vt+b$ gives
\begin{equation}
    v=0.989882\pm 0.000095 ,
\end{equation}
in agreement with the expected continuum velocity. We believe the deviation
from unity is attributable to finite-size effects, the finite lattice spacing,
and the presence of non-universal high-energy modes excited by the local quench.

To obtain a controlled approximation to the continuum CFT, all physical length
scales used in the entanglement calculations are taken to be much larger than
the lattice spacing. In particular, the lattice spacing $a$ is chosen such that
\begin{equation}
    a \ll l_A,\; l_B,\; d ,
\end{equation}
where $l_A$ and $l_B$ are the sizes of the subsystems under consideration and
$d$ is the separation between them. In this regime, nonuniversal lattice-scale
effects are suppressed, while the long-distance behavior is governed by the
$c=1$ Dirac CFT.